\documentclass[12pt]{article}

\usepackage{amssymb}
\usepackage{amsfonts}
\usepackage{amsmath}
\usepackage{amscd}
\usepackage{txfonts}
\usepackage{tikz}
\usepackage{geometry}
\usepackage{latexsym}
\usepackage{epsfig}
\usepackage{graphicx}
\usepackage{verbatim}
\usepackage{babel}
\usepackage[latin1]{inputenc}
\usepackage{fancyvrb}
\usepackage{pdfpages}
\usepackage{caption}
\usepackage[size=scriptsize]{subcaption}
\usepackage{float}
\usepackage{xcolor, colortbl}
\usepackage{hyperref}
\usepackage{placeins}
\usepackage{rotating}
\usepackage[title]{appendix}

\definecolor{gray1}{gray}{0.8}
\definecolor{gray2}{gray}{0.7}

 \numberwithin{equation}{section}

\def\min{\mathop{\rm min}\nolimits}

\def\KL{\mathop{\rm KL}\nolimits}
\def\JS{\mathop{\rm JS}\nolimits}
\def\Com{\mathop{\rm Com}\nolimits}

\def\MPF{\mathop{\rm MPF}\nolimits}
\def\MNE{\mathop{\rm MNE}\nolimits}
\def\GNE{\mathop{\rm GNE}\nolimits}
\def\MTF{\mathop{\rm MTF}\nolimits}
\def\H{\mathop{\rm H}\nolimits}
\def\NE{\mathop{\rm NE}\nolimits}
\def\Degree{\mathop{\rm Degree}\nolimits}
\def\EVC{\mathop{\rm EVC}\nolimits}
\def\PE{\mathop{\rm PE}\nolimits}

\let\frac=\dfrac
\let\leq\leqslant

\makeatletter
\providecommand{\textlangle}{$\langle\m@th$}
\providecommand{\textrangle}{$\rangle\m@th$}
\makeatother
\providecommand{\keywords}[1]
{
  \small	
  \textbf{\textit{Keywords---}} #1
}

\author{
  \small{M. L\'opez$^{1,*}$, R. Mansilla$^2$}\\
  \small{$^1$Facultad de Ciencias, Universidad Nacional Aut\'onoma de M\'exico,}\\ \small{Ciudad Universitaria, 04510, Ciudad de M\'exico, M\'exico.}\\
  \small{$^2$Centro de Investigaciones Interdisciplinarias en Ciencias y Humanidades,}\\\small{Universidad Nacional Aut\'onoma de M\'exico, Ciudad Universitaria, 04510, Ciudad de M\'exico, M\'exico.}\\
  \small{$^*$ Correspondence: mariolopper@gmail.com}
  }
\title{\textbf{Ordinal Synchronization and Typical States in High-Frequency Digital Markets}}
\date{}
\begin{document}

  \maketitle

\begin{abstract}
In this paper we study Algorithmic High-Frequency Financial Markets as dynamical networks. After an individual analysis of 24 stocks of the US market during a trading year of fully automated transactions by means of ordinal pattern series, we define an information-theoretic measure of pairwise synchronization for time series which allows us to study this subset of the US market as a dynamical network. We apply to the resulting network a couple of clustering algorithms in order to detect collective market states, characterized by their degree of centralized or descentralized synchronicity. This collective analysis has shown to reproduce, classify and explain the anomalous behavior previously observed  at the individual level. We also find two whole coherent seasons of highly centralized and descentralized synchronicity, respectively. Finally, we model these states dynamics through a simple Markov model.

\end{abstract}

\keywords{Ordinal Patterns, High-Frequency Trading, Algorithmic Trading, Networks, Synchronicity, Clustering}

\section{Introduction}

Financial markets are highly complex evolving systems, which means that their statistical dynamics is constantly redefined by the interaction of economic agents. Thus, it is not surprising that, in order to understand these dynamics, much effort have been recently dedicated to study them as dynamical networks which can be analyzed and classified, either topologically to quantify crisis periods \cite{99Mantegna} \cite{03Onnela} or by clustering them to detect market states \cite{02Marsili}\cite{12Munnix} \cite{15Chetalova} \cite{19Masuda} \cite{21Chakraborti} and posibly early precursors for the catastrophic ones \cite{20Chakraborti} \cite{18Pharasi}.

That approach looks very promising for studying Algorithmic High-Frequency Trading, whose rise during the last decades was recently shown to display an even higher degree of networked structure and complexity than those of traditional markets \cite{21Musciotto}. However, those network-based studies have been carried out mainly by defining edge weights through correlation matrixes, which amounts to ignore non-linear interactions (see \cite{17Kim} for an exception, which does not use cluster analysis, but Machine Learning techniques). This is not adequate for high-frequency data, normally expected to be very noisy and highly non-linear and non-stationary. Thus, in order to successfully apply the same network-clustering pipeline of those previous works and, at the same time, be able to detect locally complex non-linear interactions, we define dynamical networks of stocks by means of their transcript synchronicity, a measure of pairwise coupling of time series defined through ordinal patterns, a tool which has been successfully applied to discern non-linear deterministic and stochastic dynamics in real-world data \cite{12Zanin}. 

Once this has been done, we propose to study the obtained dynamical network through the distribution of its Eigenvector Centrality and Degree, two well known measures of connectivity for network nodes which we use to define suitable phase representation spaces in order to detect meaningful market states through a couple of clustering algorithms. This allows us to detect two whole coherent and quantitatively distinguishable seasons, characterized by their degree of centralized/decentralized synchronicity.

So, the goal of the paper is two fold: to adapt an increasingly popular methodology for studying financial markets as dynamical networks and clusters as market states to the needs of Algorithmic High-Frequency Trading Data, and to show with its application to a particular data set of fully automated transactions its potential for detection of collective dynamical regimes.

In order to underline the necessity of collective analysis, we include a section in which ordinal pattern analysis for individual stocks is carried out as a start point, and whose findings are latter shown to be reproduced, extended and further explained at the collective level. For this, we use some of the most common information-theoretic measures related to ordinal patterns.

The paper is organized as follows: Section 2 introduces the theoretical background necessary to apply ordinal pattern analysis. Section 3 describes the characteristics of the data. In Section 4 we apply the previosly defined measures to individual stocks, in order to detect anomalous behaviors. Section 5 contains the definition of our transcript synchronicity coefficient, which we use to define our dynamical network and discuss the more convenient phase representation spaces for the sake of our analysis. In Section 6 we carry out the clustering analysis to detect typical market states, which in Section 7 are modeled as first order Markov processes. Section 8 contains the conclusions.

%%%%%%%%%%%%%%%%%%%%%%%%%%%%%%%%%%%%%%%%%%
\section{Theoretical Background: Ordinal Patterns}\label{sec: theory}

Permutation entropy was introduced in \cite{02Bandt} as a (non-parametric) complexity measure, robust to dynamical noise, invariant with respect to nonlinear monotonous transformation and computationally efficient. It is defined as follows: First, given a time series $x_n$ for $n = 1,\dots,N$ and two parameters $m<N$ and $l$, the pattern length and the time lag respectively, we consider, $S_m$, the group of permutations of length $m$ and, for $0 < t \leq N-(m - 1)\cdot l$, we say that the sliding window $x^l_m(t) = (x_t, x_{t+l},\dots,x_{t+(m - 1)\cdot l})$ is of type $\pi_t\in S_m$ if $\pi_t = (i_1,\dots,i_m)$ is the only $m$-permutation satisfying the following conditions:
\begin{enumerate}
\item[(1)] $x_{t+i_s\cdot l}\leq x_{t+i_{s+1}\cdot l}$ for $s = 1,\dots,m-1$, and 
\item[(2)] $i_s < i_{s+1}$ if $x_{t + i_s\cdot l} = x_{t + i_{s+1}\cdot l}$,
\end{enumerate}

and we denote this with $\Phi(x^l_m(t)) = \pi_t$. 

For example: if $x = (1.2, 3.2, 2.3, 1.4, 1.1, 4.3, 3.1)$ is a time series then, since $x_5 < x_1 < x_4 < x_3 < x_7 < x_2 < x_6$, we have the 5-length ordinal patterns:
\[\begin{array}{c}
\pi_1 = \Phi(x^1_5(1)) = (4, 0, 3, 2, 1)\\
\pi_2 = \Phi(x^1_5(2)) = (3, 2, 1, 0, 4)\\
\pi_3 = \Phi(x^1_5(3)) = (2, 1, 0, 4, 3),\\
\end{array}\]
thus obtaining the 5-length ordinal pattern series
\[
\pi_x = ((4, 0, 3, 2, 1), (3, 2, 1, 0, 4), (2, 1, 0, 4, 3)),  
\]
and similarly from the time series $y = (3.2, 4.2, 5.1, 0.4, 0.9, 2.3, 3.4)$ we obtain 
\[
\pi_y = ((3, 4, 0, 1, 2), (2, 3, 4, 0, 1), (1, 2, 3, 4, 0)).
\]

The study of this new ordinal patterns series $\{\pi_t\}_{t<N-(m-1)\cdot l}$, has been applied to biomedicine \cite{12Cysarz} \cite{12Parlitz} \cite{11Nicolaou} \cite{09Ouyang} \cite{12Faes} \cite{12Bian} \cite{10Li} \cite{08Olofsen}, paleoclimatology \cite{18Garland}, economics \cite{12Ruiz} \cite{08Zanin} \cite{09Zunino} \cite{10Zunino} \cite{12Zunino} \cite{11Zunino}, geology \cite{19Pessa} and engineering \cite{12Yan} for classification and prediction of deterministic and stochastic non-linear dynamics.

The choice for $m$ is generally unproblematic, as it is understood that $m$ must be as large as posible without compromising statistical reliability when measuring information-theoretic quantities on $m$-length ordinal patterns distributions (to be definied below), for which is enough to set $m! << N$, being $m!$ the cardinality of the permutation group $S_m$ \cite{08Matilla} \cite{18Garland}. Our data (see next section) dictates the choice $m = 5$. As for the lag time $l$, it has been shown to be critically important, since a na\"{\i}ve choice could lead to spurious results, making thus necessary to carry out a multiscale analysis (that is, varying $l$) before drawing any conclusions \cite{20Olivares_1}. Our results, however, have shown to be highly independent of this parameter, and our conclusions are the same no matter which value of $l$ we pick, in a range going from $l=1$ to $l=100$ (see \autoref{appendix: multiscale} for some figures supporting this claim). This is in itself an interesting result: we are observing here a phenomenon which is present in a wide range of time scales, which allows us to drop the $l$ parameter in the subsequent discussion. The figures in the paper correspond to $l = 1$.

So, for any $\pi\in S_m$ its relative frequency is defined as:
\[
p_m(\pi)=\frac{\textrm{\#}\{t\,|\, t\leq N-m+1, \Phi(x_m(t)) = \pi \} }{N-m+1}.
\]

Although the original proposal by Bandt and Pompe consisted basically in the study of time series through Shannon entropy of the ordinal patterns distribution, called \textit{permutation entropy} (PE) and given by
\[
\PE(m) = -\sum_{\pi\in S_m}p_m(\pi)\log p_m(\pi),
\]

a vast stream of theoretical and methodological approaches has developed since then, giving place to a constellation of complexity measures, important examples of which are weighted permutation entropy \cite{13Fadlallah} \cite{09Liu}, symbolic transfer entropy \cite{08Staniek} \cite{16Amigo}, statistical complexity \cite{10Zunino} \cite{95Lopez} \cite{04Lamberti}, various measures of coupling and synchronicity \cite{10Li} \cite{08Bahraminasab}\cite{16Amigo} \cite{09Monetti} and network-based measures \cite{19Pessa} (we will define some of them below). Other studies had combined these analytical tools with Machine Learning techniques \cite{11Nicolaou} \cite{17Keller}, and still others had exploited the algrebraic structure of $S_m$ \cite{09Monetti} \cite{12Amigo} \cite{13Bunk} \cite{16Amigo}.

We must be very careful in applying permutation entropy, the authors of \cite{17Zunino} warn us. The presence of equal consecutive values in the original time series, which is frequently dealt with by preserving the temporal order in the corresponding permutation, could lead to draw false conclusions by detecting spurious patters. To adress this issue one can sum low amplitude noise to the original series in order to (randomly) break equalities, as proposed in \cite{15Quintero}, and that is exactly what we do in this work, with an artificially generated uniform distribution series of amplitude $10^{-7}$.

We will need some (more or less) classic definitions for the next sections. Given two probability distributions $p, q$ defined on a finite set, its Jensen-Shannon divergence is defined as
\[
D_{\JS} (p, q) = \frac{D_{\KL}(p\,\vert\vert\, M) + D_{\KL}(q\,\vert\vert \,M)}{2}, 
\]
where $M = (p+q)/2$ and $D_{\KL}$ is the Kullback-Leiber divergence:
\[
D_{\KL}(p\,\vert\vert \,q) = \sum_i p_i\log\frac{p_i}{q_i},
\]

or the relative entropy from $q$ to $p$, of which $D_{\JS}$ is thus a smoothed, symmetric version.

So, if $p_m$ is the probability distribution of ordinal patterns of the time series $x_i$, its \textit{statistical complexity} (Com) is defined as \cite{10Zunino} \cite{95Lopez} \cite{04Lamberti}
\[
\Com(p_m) = D_{\JS}(p_m, u_m)\,\H(p_m), 
\]
where $u_m$ is the uniform distribution on $S_m$ and $\H(\,\cdot\,)$ is Shannon entropy. Thus, $\Com(p)$ aims to measure complexity through a trade-off between randomness and determinism: while Shannon entropy increases as $p_m$ aparts away from determinism towards randomness, reaching a maximum for $u_m$, its divergence from $u_m$ grows as $p_m$ aparts away from randomness. $\PE(m)$ and $\Com(p)$ have been jointly used to classify dynamical regimes \cite{10Zunino}.

Missing Patterns Frequency (MPF) \cite{07Amigo} is defined as 
\[
\MPF(p_m) = \frac{\textrm{\#}\{\pi \in S_m \,|\, p_m(\pi) = 0\}}{m!}.
\]

As deterministic dynamics are expected to display a relatively small set of ordinal patterns in contrast to random dynamics \cite{07Amigo} \cite{08Zanin} \cite{09Zunino}, MPF is understood to quantify determinism degrees.

Yet another methodology for studying ordinal patterns, this time as nodes of a network, has been proposed in \cite{17McCullogh}\cite{11Campanharo} \cite{17Zhang} \cite{19Pessa}, where the directed edges are weighted according to the transition probability of passing from one pattern to another inmediatly in time, thus taking patterns as states of the process. So, for the ordinal sequence $\{\pi_t\}_{t<N-m+1}$ nodes are defined as the ordinal patterns and the weight of their edges are given by the transition probabilities $p_m(\pi'|\,\pi)$ of observing for some $t$ that $\pi_{t+m} = \pi'$ given that $\pi_t = \pi$. We can then compute the node entropy as 
\[
\NE(\pi) = -\sum_{\pi'\in S_m} p_m(\pi'|\,\pi)\log(p_m(\pi'|\,\pi)).
\]

These measures, unlike the previously defined, aim to quantify determinism not in terms of pattern frequency, but of pattern transitions in time. Minimum Node Entropy (MNE) and Global Node Entropy (GNE) are defined as \cite{20Olivares} 
\[
\MNE(p_m) = \min\{\NE(\pi) \,|\,\pi\in S_m \}\,\,\textrm{and}\,\,\GNE = \sum_{\pi \in S_m}p_m(\pi)\NE(\pi), 
\]
respectively, so MNE measures how deterministic can a pattern be in a given network, while GNE gives us a global, weighted score of pattern transitions determinism. Finally, missing transitions frequency (MTF) is defined as 
\[
\MTF(p_m) = \frac{\textrm{\#}\{(\pi,\pi') \in S_m\times S_m \,|\, p_m(\pi'|\,\pi) = 0\}}{(m!)^2},
\]
and is of course the analogous measure of MPF for pattern transitions.

A classification plane is proposed in \cite{20Olivares}, whose axes are given by GNE and the MNE, both of them measured using non-overlapping ordinal patterns to avoid transition constrains. This methodology seems to be very useful, not to mention intuitive, to distinguish between (linear) stochastic and (non-linear) deterministic dynamics. When applied to stock market time series, this methodology allows the authors to discriminate financial dynamics from fractional Gaussian noise, since financial scores lie below the diagonal around which the noises cluster. We will find this plane useful in the following discussion.

%%%%%%%%%%%%%%%%%%%%%%%%%%%%%%%%%%%%%%%%%%
\section{Data}
The data used in this investigation are the time series of prices of the automated (algorithmic) operations that occurred from March 7, 2018 to March 7, 2019 in the US market, with a total of 251 trading days and 539,834,024 records.

  \begin{table}[!ht]
  \centering
  \begin{tabular}{c c c}
  \rowcolor{gray2}
  & Market code & Name \\
  \rowcolor{gray1}
  1 & ABT & Abbott Laboratories \\
  \rowcolor{gray2}
  2 & BAC & Bank of America Corporation \\
  \rowcolor{gray1}
  3 & BMY & Bristol-Myers Squibb Company\\
  \rowcolor{gray2}
  4 & C & Citigroup Inc.\\
  \rowcolor{gray1}
  5 & CSCO & Cisco Systems, Inc.\\
  \rowcolor{gray2}
  6 & F & Ford Motor Company\\
  \rowcolor{gray1}
  7 & FB & Facebook, Inc. \\
  \rowcolor{gray2}
  8 & FOXA & Tweenty-First Century Fox, Inc.\\
  \rowcolor{gray1}
  9 & GE & General Electric Company \\
  \rowcolor{gray2}
  10 & GM & General Motors Company\\
  \rowcolor{gray1}
  11 & HPQ & HP Inc.\\
  \rowcolor{gray2}
  12 & INTC & Intel Corporation\\
  \rowcolor{gray1}
  13 & KO & The Coca-Cola Company \\
  \rowcolor{gray2}
  14 & MDLZ & Mondelez International, Inc.\\
  \rowcolor{gray1}
  15 & MO & Altria Group, Inc.\\
  \rowcolor{gray2}
  16 & MS & Morgan Stanley\\
  \rowcolor{gray1}
  17 & MSFT & Microsoft Corporation\\
  \rowcolor{gray2}
  18 & ORCL & Oracle Corporation \\
  \rowcolor{gray1}
  19 & PFE & Pfizer Inc. \\
  \rowcolor{gray2}
  20 & T & AT\&T Inc.\\
  \rowcolor{gray1}
  21 & TWTR & Twitter, Inc.\\
  \rowcolor{gray2}
  22 & USB & U.S. Bancorp\\
  \rowcolor{gray1}
  23 & VZ & Verizon Communication, Inc.\\
  \rowcolor{gray2}
  24 & WFC & Wells Fargo \& Company\\
  \end{tabular}
  
  \caption{Assets of the US market}\label{tab:stocks_US}
  \end{table}

% \begin{figure}[h!]
%   \centering
%     \includegraphics[width=0.5\columnwidth]{Stock_names_US.jpg}
% \caption{Codes and Names of the Studied Assets}\label{fig:tab_stocks_US}

% \end{figure}

The data belongs to 24 assets from the US market: ABT, BAC, BMY, C, CSCO, F, FB, FOXA, GE, GM, HPQ, INTC, KO, MDLZ, MO, MS, MSFT, ORCL, PFE, TWTR, T, USB, WFC, VZ. The details can be seen in table ~\ref{tab:stocks_US}. Our work is restricted to these 24 stocks because this is the data currently available to the authors.

The following analyses are carried out for the series of logarithmic returns \[r(t,\tau) = x(t+1,\tau)-x(t,\tau),\] where $x(t,\tau)$ is the $t$-th term of the series of means of $\tau$ seconds of logarithms of the prices of a given asset. In this work, we use $\tau = 5$ and study daily subseries to analyze daily dynamics throughout the year.

Note that it is only because our data comes from fully automated transactions that we can test our methodological insights specifically for Algorithmic High-Frequency markets, which is one of the main incentives for our inquiry, as stated in the Introduction.

%%%%%%%%%%%%%%%%%%%%%%%%%%%%%%%%%%%%%%%%%%
\section{Individual Analysis of Stocks through Ordinal Patterns}

Before addressing the analysis of collective behaviors in our data set, which is the main concern of this paper, lets take a look at the stocks individually. To do this we calculate, for the daily series of five seconds average logarithmic returns of each stock, PE, Com, MPF, MNE, GNE and MTF (the last three without overlapping patterns, just as in \cite{20Olivares}) with $m = 5$, that is, we study $5$-length ordinal patterns of daily series of returns of $5$-seconds averages of logarithms of prices. For the discussion on the choice of $m$, see \autoref{sec: theory}. Next, we plot the evolution of such quantities along the year, as well as the MNE vs GNE plane.

\begin{figure}[h!]
  \centering
    \includegraphics[width=\columnwidth]{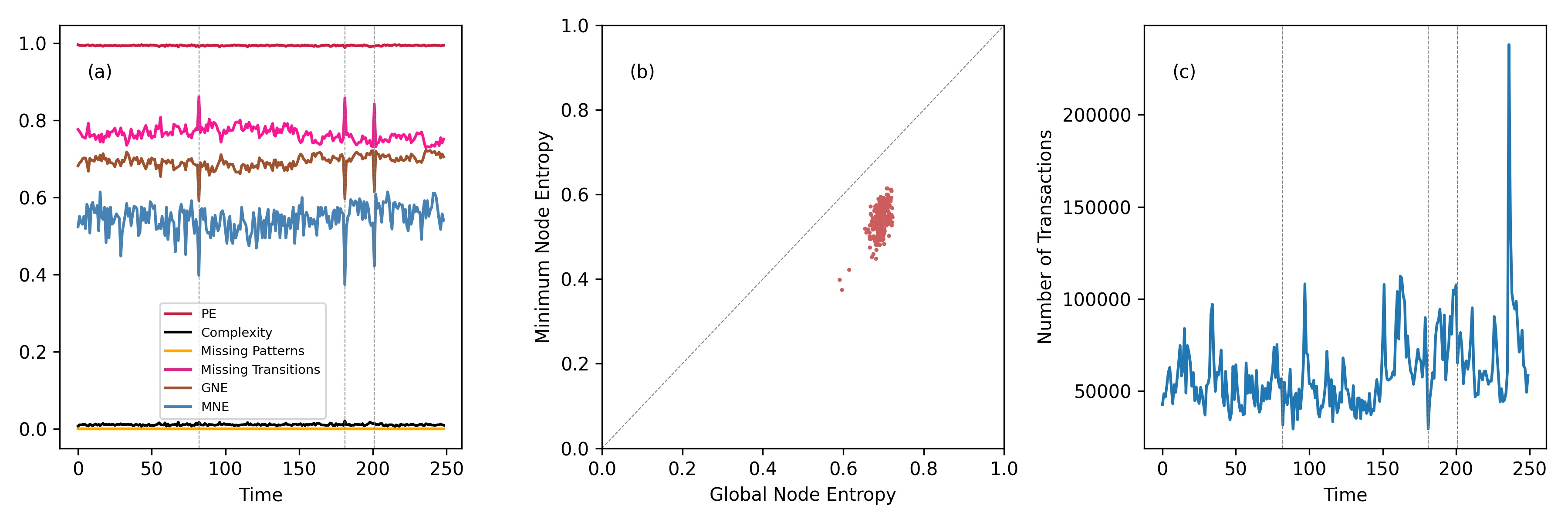}
    \caption{Ordinal Entropy Measures for KO. (a) Permutation Entropy, Complexity, Missing Pattern Frequency, Missing Transition Frequency, Global Node Entropy, Minimum Node Entropy; (b) Minimum Node Entropy vs Global Node Entropy plane; (c) Daily number of transactions.}\label{fig:KO_ent}

\end{figure}

\begin{figure}[h!]
  \centering
    \includegraphics[width=\columnwidth]{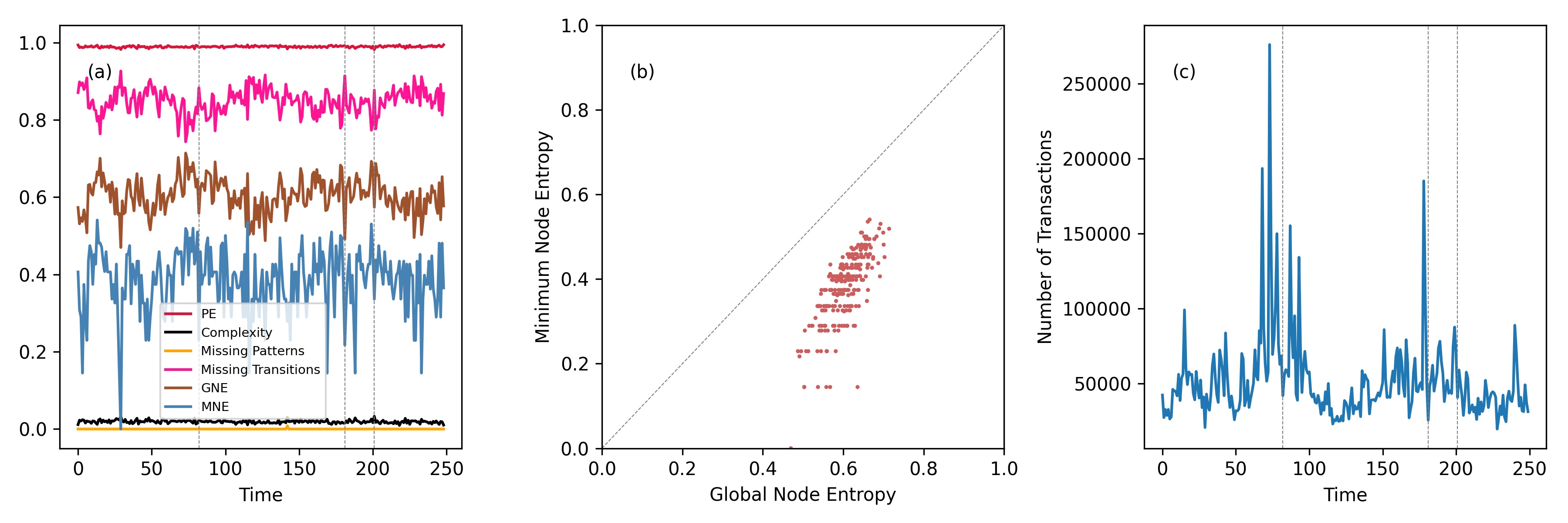}
\caption{Ordinal Entropy Measures for FOXA. (a) Permutation Entropy, Complexity, Missing Pattern Frequency, Missing Transition Frequency, Global Node Entropy, Minimum Node Entropy; (b) Minimum Node Entropy vs Global Node Entropy plane; c) Daily number of transactions.}\label{fig:FOXA_ent}

\end{figure}

For the vast majority of stocks, MNE, GNE and MTF clearly present an unusual (outlier) behavior for the days 82, 181 and 201, which we will call outlier days, while we will refer to the other days as typical (figure ~\ref{fig:KO_ent} shows this for KO; outlier days are indicated by vertical gray lines in the left and right panels). MNE and GNE abruptly decrease for the outlier days, while MTF does the opposite, thus supporting the idea that these days present complex, semi-deterministic behavior: given a certain pattern, the transition to the next one has candidates much more probable than others. In some of the stocks we can further visualize what appears to be a small number of discrete levels for MNE (see second panel in figure ~\ref{fig:FOXA_ent} for an example). This suggests the existence of a finite number of typical market states, each corresponding to a specific level of MNE. The problem with these measures is that they need length series $n >> (m!)^2$, and, as already mentioned, we could be undersampling for $m = 5$, when we begin to classify outlier days. PE, Com and MPF are not capable of detecting (visually at least) the outlier days.

By inspecting in the same figure the graphs of the daily number of transactions of each stock (right panel), we can notice that the first two outlier days are low-liquidity days for many of them, although not necessarily global minima or even specially severe drops in some others, while the last outlier does not present this behavior. Thus, low liquidity is not enough to explain this phenomenon. 

%%%%%%%%%%%%%%%%%%%%%%%%%%%%%%%%%%%%%%%%%%
\section{Collective Analysis of Stocks through Transcript Synchronicity Dynamical Networks}

To get a deeper insight on this behavior and figure out if it is, as it seems to be, a collective behavior, we study the synchronization index given by the transcript entropy of pairs of stocks, that is: for each pair $(i, j)$ of stocks and each trading day $T$, we obtain their daily ordinal pattern sequences $(\pi_i(t))^n_{t = 1}$ and $(\pi_j(t))^n_{t = 1}$ to obtain the transcript series \cite{09Monetti} as $\tau_{i,j}(t) = \pi_j(t)\circ\pi_i(t)^{-1}$, where the product and the inverse are those of the permutation group $S_m$: for $\pi, \rho\in S_m$, $\pi = (\pi_1, \dots, \pi_m)$ and $\rho = (\rho_1, \dots, \rho_m)$, 

\[
\pi \circ \rho = (\pi_{\rho_1},\dots, \pi_{\rho_m}) \textrm{, and}
\]  
\[
\pi^{-1}: \pi_k\to k \textrm{ for } k<m
\]
is the sorting operation. In this way, the transcript $\tau_{i,j}(t)$ is the result of ordering $\pi_j(t)$ according to the ordinal type of $\pi_i(t)$.

Lets illustrate this with an example: if, as in Section 2, $x = (1.2, 3.2, 2.3, 1.4, 1.1, 4.3, 3.1)$ and $y = (3.2, 4.2, 5.1, 0.4, 0.9, 2.3, 3.4)$ are two time series then, as we have seen, their respective 5-length ordinal pattern sequences are 

\[
\pi_x = ((4, 0, 3, 2, 1), (3, 2, 1, 0, 4), (2, 1, 0, 4, 3))
\]
and
\[
\pi_y = ((3, 4, 0, 1, 2), (2, 3, 4, 0, 1), (1, 2, 3, 4, 0)).
\]

The group inverse of $\pi_x(1) = (4, 0, 3, 2, 1)$, which denotes the function 

\[\begin{array}{c}
0\to 4\\  1\to 0\\  2\to 3\\  3\to 2\\  4\to 1,
\end{array}\] 

is just the inverse function

\[\begin{array}{c}
4\to 0\\ 0\to 1\\ 3\to 2\\ 2\to 3\\ 1\to 4,
\end{array}\]
or, conveniently, expresed, $\pi_x(1)^{-1} = (1, 4, 3, 2, 0)$. After similar calculations for the remaining ordinal patterns we obtain the series of group inverses 

\[
\pi_{x^{-1}} = ((1, 4, 3, 2, 0), (3, 2, 1, 0, 4), (2, 1, 0, 4, 3)).
\]

Now, to obtain our first transcript we compose the two functions $\pi_x(1)^{-1} = (1, 4, 3, 2, 0)$ and $\pi_y(1) = (3, 4, 0, 1, 2)$, thus obtaining the function
\[\begin{array}{ccccc}
  & \pi_x(1)^{-1} & & \pi_y(1) & \\
0 & \to & 1 & \to & 4\\
1 & \to & 4 & \to & 2\\
2 & \to & 3 & \to & 1\\
3 & \to & 2 & \to & 0\\
4 & \to & 0 & \to & 3  
\end{array}\]

better represented as $\pi_y(1)\circ \pi_x(1)^{-1} = (4, 2, 1, 0, 3)$. After repeating the operations with the other pairs of ordinal patterns, we finally get the transcript series:
\[
\tau_{x, y} = ((4, 2, 1, 0, 3), (0, 4, 3, 2, 1), (3, 2, 1, 0, 4))
\]

We take the normalized entropy of this transcript series $H^{transcript}_T(i, j)$ (that is, the usual permutation entropy of this transcript series) as a (daily) coefficient of desynchronization, and one minus that quantity as our measure of synchronization, which we will call transcript synchronization. Transcript synchronization measures the diversity of transcripts: low transcript synchronization means high variety of transcripts, that is, a lot of different transcripts and then a lot of information is needed to deduce the (ordinal) dynamics of one series given complete knowledge of the other, and analogously for high transcript synchronization. Transcripts have been applied to study synchronization in time series in \cite{12Amigo}, \cite{13Bunk}, \cite{13Monetti_1}, \cite{16Amigo}.

Thus we obtain for each trading day a (symmetric) transcript synchronization matrix $H^{transcript}_T$ of dimension $n_{stock} \times n_{stock}$ whose $ij$ value is the transcript synchronization of stocks $i$ and $j$ during day $T$. By considering each of these daily matrixes as adjacency matrixes, we obtain a dynamical weighted network through the year, the nodes of which are the stocks and the weight of whose edges are given by our transcript synchronization coefficient. We can thus analyze our time series with classical network-based measures, following \cite{02Marsili} \cite{12Munnix} \cite{15Chetalova} \cite{18Pharasi} \cite{19Masuda} \cite{99Mantegna} \cite{20Chakraborti} \cite{21Chakraborti}, which have done that for correlation matrixes, and \cite{17Kim}, where mutual information networks are studied. We consider two well known such network measures here: degree and eigenvector centrality.

Given a stock labeled as $i$ and a trading day $T$, we define its degree as 
\[
\Degree(i, T) = \frac{1}{C}\sum^{n_{stocks}}_{j =1,\, j\neq i} H^{transcript}_T(i, j),
\]
so the degree of a node is just the sum of its transcript synchronization with all the other stocks and the normalizing constant $C$ is such that $\sum_i\Degree(i, T) = 1$, while its eigenvector centrality is defined as the $i$-th component of the normalized solution to the equation
\[
\EVC(i, T) = \frac{1}{\lambda}\sum^{n_{stocks}}_{j=1}H^{transcript}_T(i, j)\EVC(j, T)
,\]
where $\lambda$ is the largest eigenvalue of the adjacency matrix $H^{transcript}_T$. This is equivalent to find the normalized eigenvector with positive components corresponding to $\lambda$, which is known to exist by the Frobenius Theorem. So for each trading day $T$ we have Degree and EVC, two $n_{stocks}$-dimensional vectors widely applied as measures of the importance, centrality or connectedness of a node in the network.

\begin{figure}[h!]
  \centering
    \includegraphics[width=0.48\columnwidth]{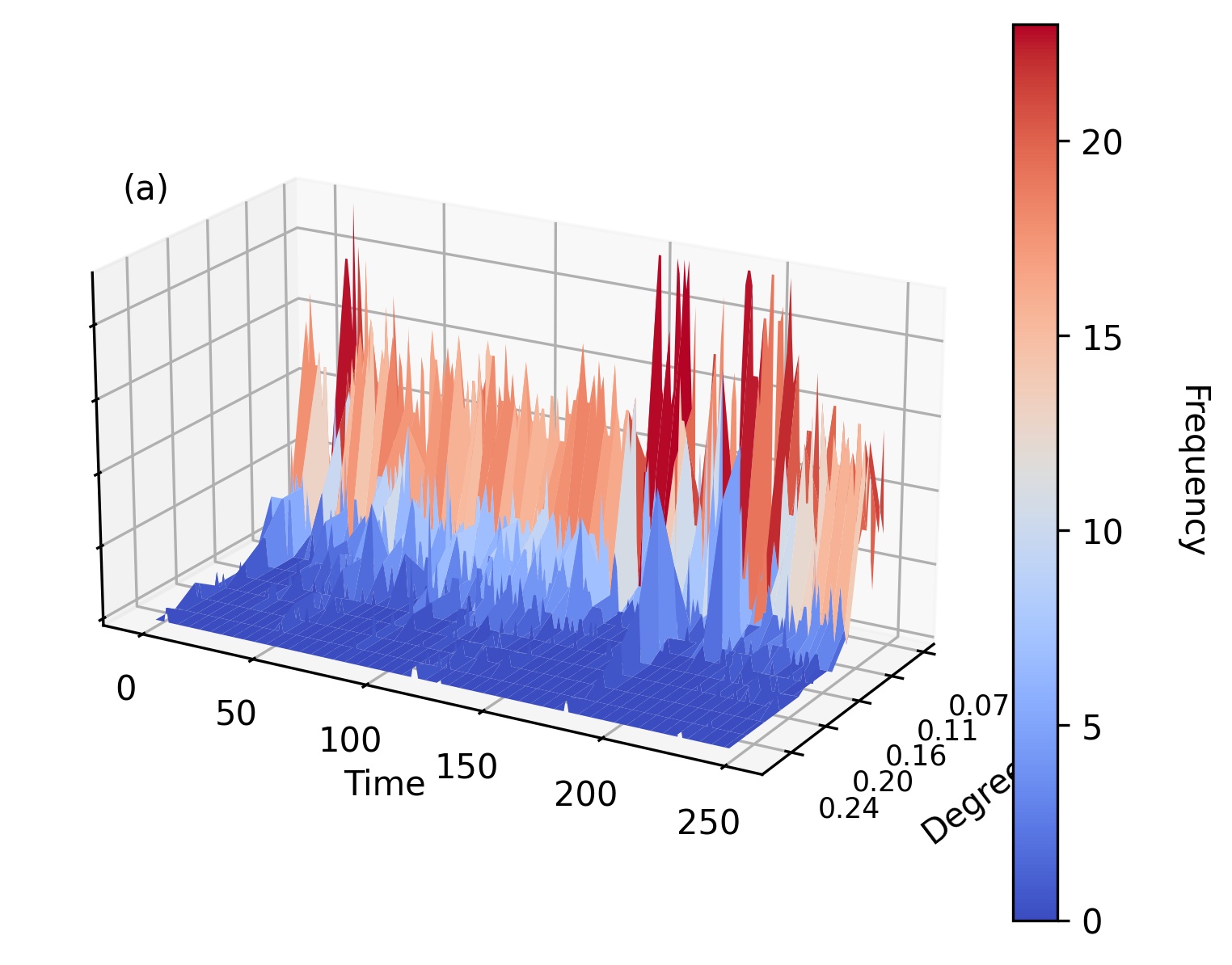}
    \includegraphics[width=0.48\columnwidth]{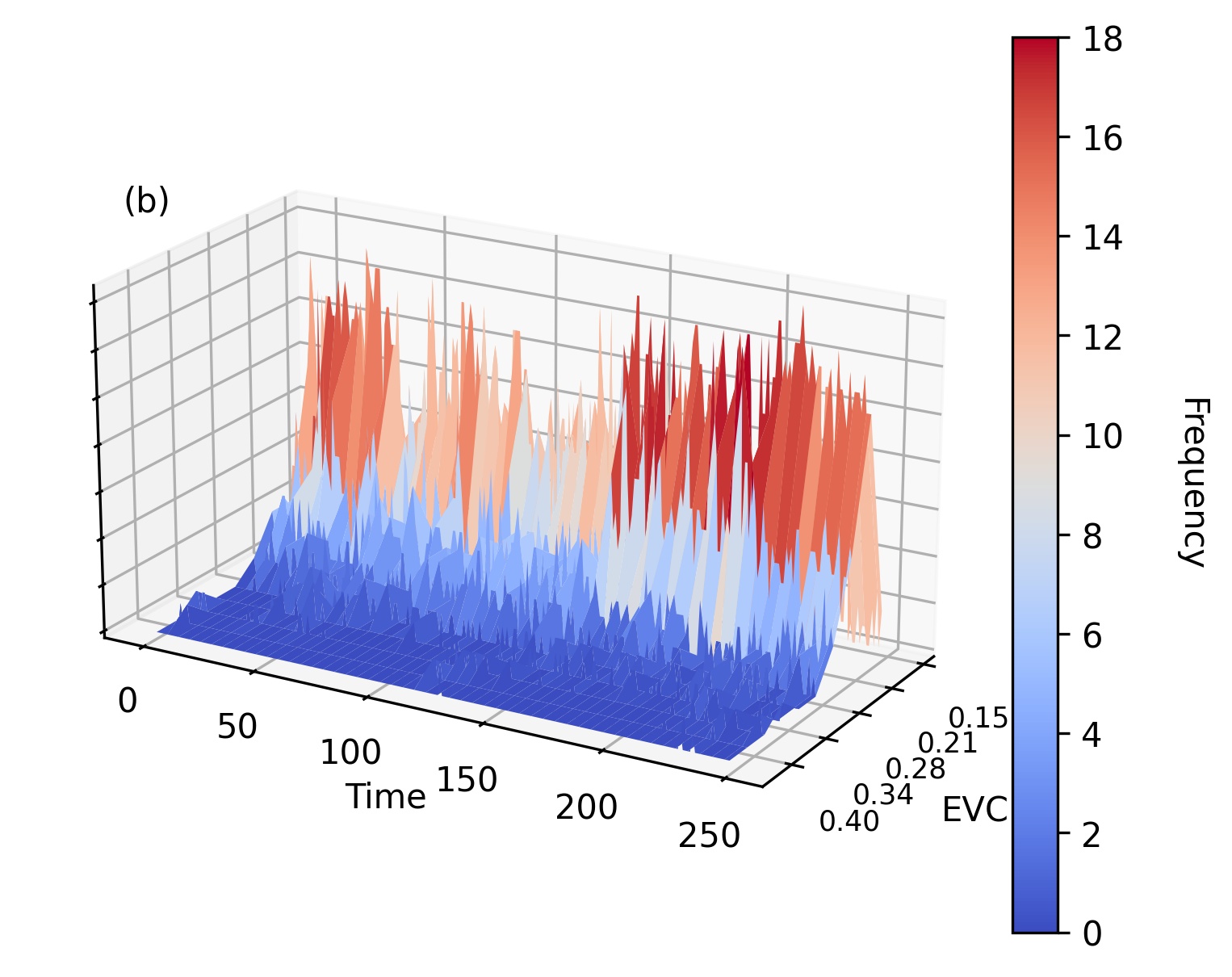}    
\caption{Evolution Through the Trading Year of Histograms of (a) Degree and (b) Eigenvector Centrality for Transcript Synchronization Dynamical Network.}\label{fig:dynamical_hist_transcript}

\end{figure}

\begin{figure}[h!]
  \centering
    \includegraphics[width=\columnwidth]{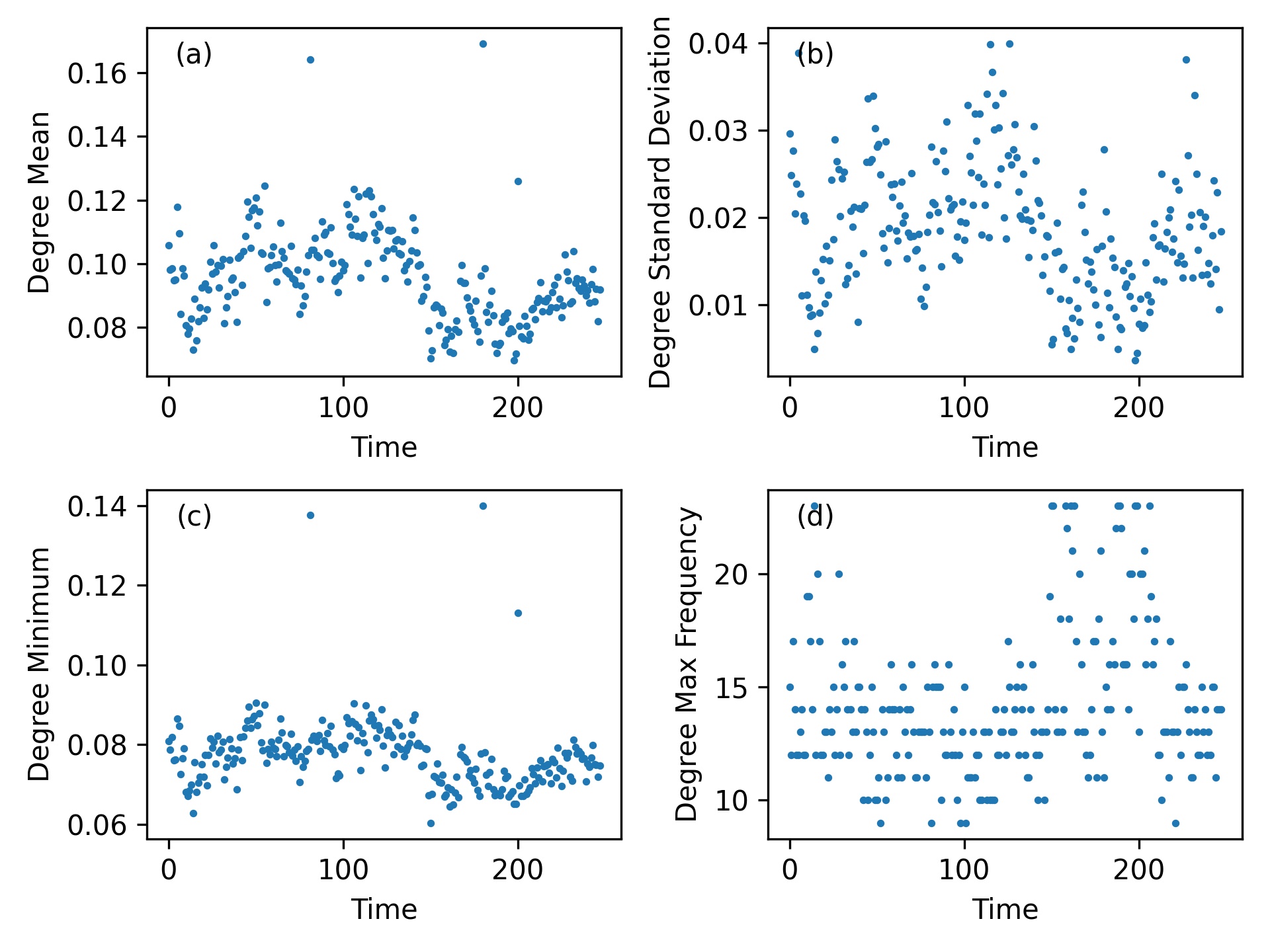}
\caption{Evolution of Degree (a) Mean, (b) Standard Deviation, (c) Minimun and (d) Maximum.}\label{fig:degree_measures}

\end{figure}

In what follows we remove the first day of the year, for it turns out to be a very peculiar outlier. If we plot the evolution of the histograms of Degree and EVC (figure ~\ref{fig:dynamical_hist_transcript}) we can observe a very important increment in the mode of Degree (this will be much more clear in figure ~\ref{fig:degree_measures}) during the outlier days and further detect not just those three days, but what appears as two complete consecutive, although intermitent, regimes of highly centralized and decentralized connectivity, the latter weakly present at the beginning of the year and again with much more persistence roughly from day 150 to day 210, and the former just between those periods, approximately from day 50 to day 150. Thus, our outlier days seem to be an extreme manifestation of these collective dynamics. Recall that the detection of outlier days and different states at this collective level was by no means an obvious thing to expect, since the only measures capable of detecting them in the individual level were those measuring transitions between consecutive ordinal patterns, while our transcript synchronization coefficient focuses only in simultaneous pairs of patterns across the market.

\begin{figure}[h!]
\centering
\includegraphics[width=\columnwidth]{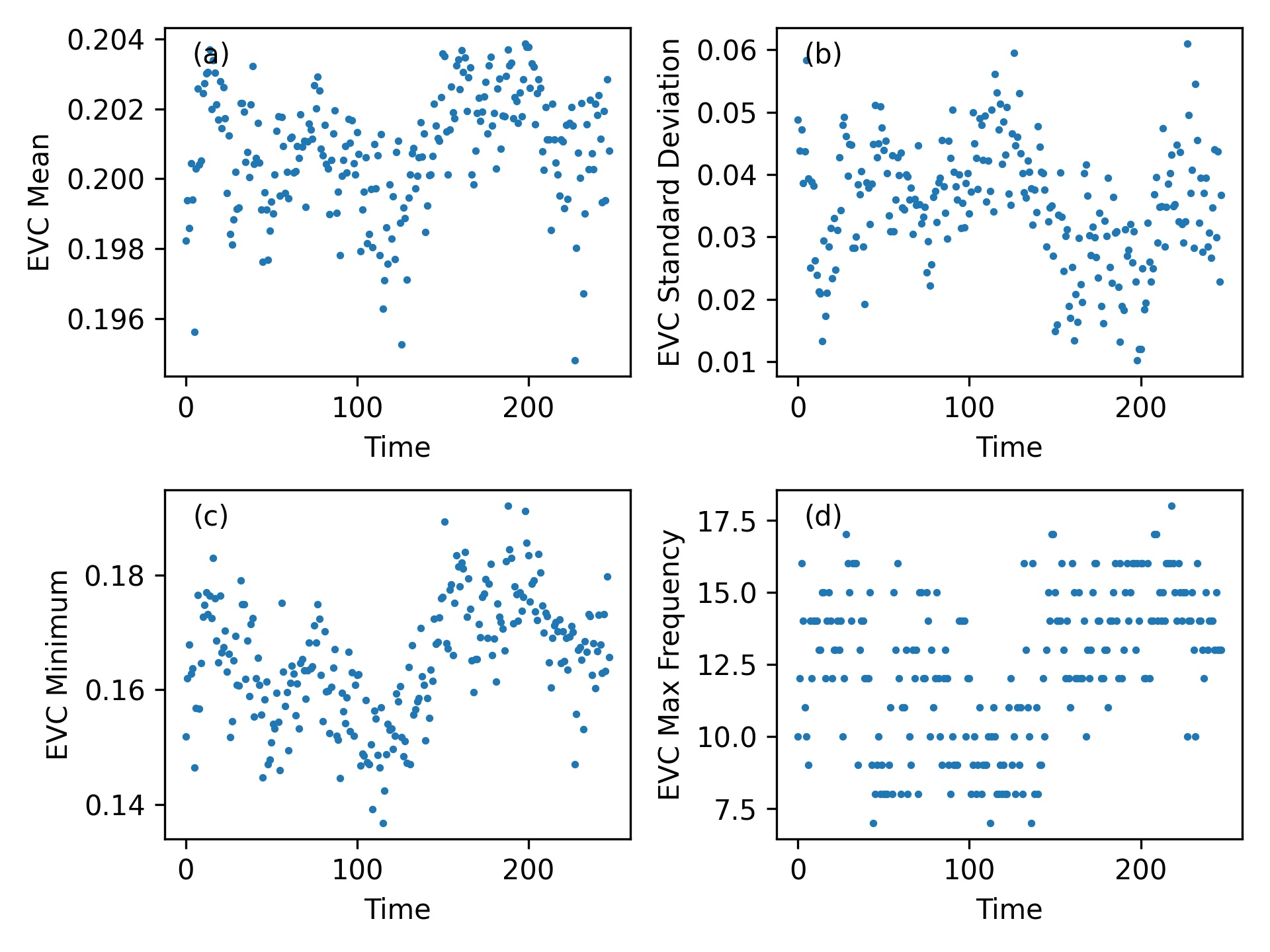}
\caption{Evolution of Eigenvector Centrality (a) Mean, (b) Standard Deviation, (c) Minimun and (d) Maximum.}\label{fig:eigen_measures}
\end{figure}

To further illustrate this behavior, in figures ~\ref{fig:degree_measures} and ~\ref{fig:eigen_measures} we plot mean, standard deviation, minimum value and maximum frequency of the previously plotted daily histograms. First of all, for Degree one can easily confirm the presence of the outlier days, as its mean and minimum abruptly increase. Also, and this holds for both figures, at the beginning of the year (March) and during the highly decentralized season we can observe that the mean of the distributions of EVC and Degree increases, while standard deviation decrease, that is to say, the histograms shrink. 

Recall that here highly decentralized synchronization means a more uniform distribution of degree-eigenvector centrality among stocks, that is, most of the nodes are equally important in the network. That's why high connectedness can be associated with the shrinking of histograms, in opposition to centralized synchronicity, which happens when there are clearly dominating stocks, much more centrals to the network than the others.

\begin{figure}[h!]
  \centering
    \includegraphics[width=\columnwidth]{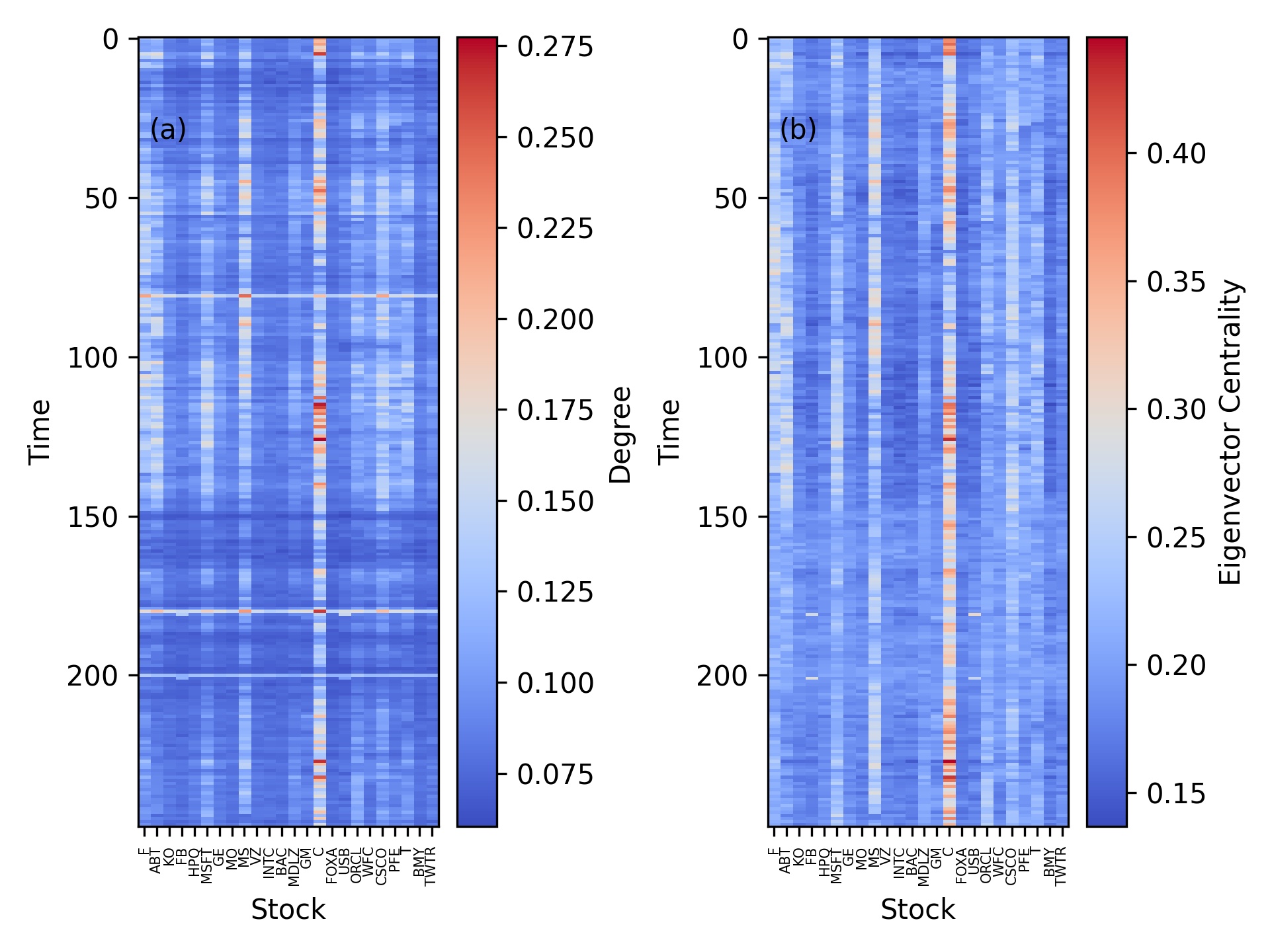}
\caption{Evolution Through the Trading Year of (a) Degree and (b) Eigenvector Centrality per stock.}\label{fig:dynamical_deg_eigen}

\end{figure}

Also in the evolution of Degree and EVC per stock (figure ~\ref{fig:dynamical_deg_eigen}), we can see a different regime in about the same period, when their values seem to be distributed more uniformly between stocks than before: both high and low scored stocks tend to equalize each other towards a intermediate value (in the heat graph, blues and reds moves towards white). An exceptional stock, that seems to remain very central to the network through the entire year but with particular force during the highly centralized season is Citigroup (C). The next stock in importance seems to be Morgan Stanley (MS), also from the Financial sector. Interestingly, Facebook (FB) and U. S. Bancorp (USB) increase their influential score just the day after the last two oulier days. Degree very clearly shows again the presence of outlier days, and we can spot the stocks driving their dynamics: the already noted Citigroup and Morgan Stanley, but also Cisco Systems (CSCO), Oracle (ORCL), Microsoft (MSFT) and Ford (F): with the exception of Ford, all of them belonging to the digital technology sector. The last outlier day seems to be a very decentralized one, its corresponding row displaying a remarkably uniform color.

Of course, the visual evidence here presented is not clear enough as to conclude in any formal way the existence of such dynamical regimes. It is to formalize and further investigate this structure that the next sections are dedicated.

Let's say that if the normalized eigenvector centrality or degree vectors are understood to be measures of market direction and strength, by measuring which specific stocks are driving it and how much, their histograms, by disregarding the latter information, can be understood to measure just the intensity of this ``market force", how much stocks are well connected and so on.

%%%%%%%%%%%%%%%%%%%%%%%%%%%%%%%%%%%%%%%%%%
\section{Clustering Analysis}
Next, we want to apply a couple of clustering algorithms to our dynamical network, looking for typical market states, for which we need first to define a matrix distance $\zeta(A, B)$ to measure similarity between daily transcript synchronization networks. After inspection of similarity matrixes with different metrics ($L^1$, $L^2$, Jensen-Shannon Distance (the square root of $D_{\JS}$)) and different phase representation (distances are measured between the whole transcript synchronization matrixes, the EVC or degree vectors, and also between their histograms) one can conclude that whether we use $L^2$ or Jensen-Shannon Distance we arrive to similar results, and the difference lies in the phase representations, of which the histograms of degree or/and EVC seem to be the most informative ones. Thus, in what follows we will be studying not the adjacency matrixes themselves, but the histograms of their EVC and Degree vectors. Since it is more meaningful than $L^2$ norm when measuring distances between probability distributions (it can be understood as a minimum redundancy measure \cite{03Endres}), we will use the Jensen-Shannon distance for the dendogram clustering. This is done also in order to reflect the structure found in the previous section.

\begin{figure}[h!]
  \centering
    \includegraphics[width=\columnwidth]{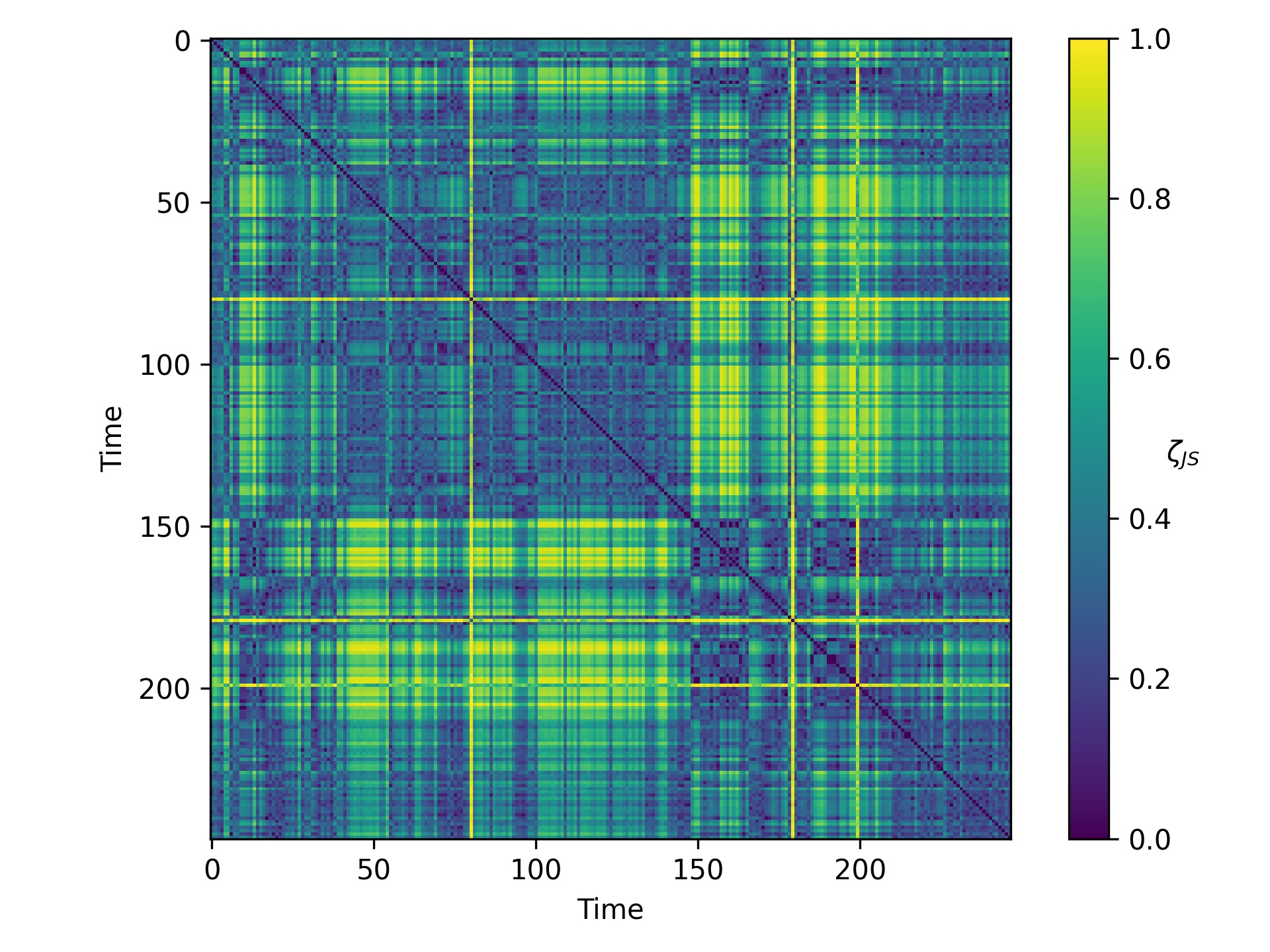}
\caption{Jensen-Shannon Distance Matrix in Degree Phase Space. Its $ij$ term equals the Jensen-Shannon distance between networks corresponding to days $i$ and $j$}\label{fig:sim_time_matrix_Deg}

\end{figure}

\begin{figure}[h!]
  \centering
    \includegraphics[width=\columnwidth]{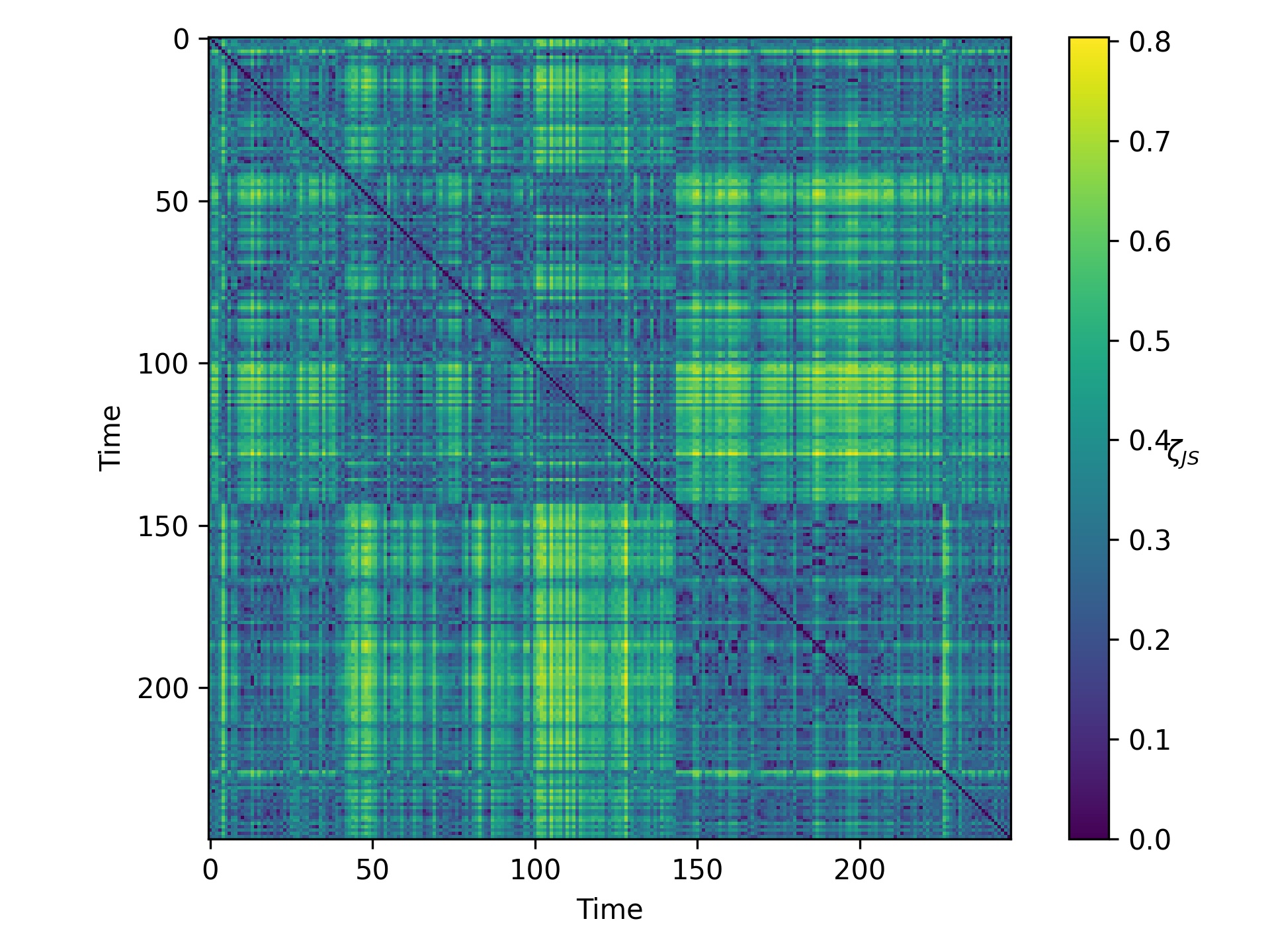}
\caption{Jensen-Shannon Distance Matrix in EVC Phase Space. Its $ij$ term equals the Jensen-Shannon distance between networks corresponding to days $i$ and $j$}\label{fig:sim_time_matrix_EVC}

\end{figure}

If we then plot the Jensen-Shannon distance matrix $D_{ij} = \zeta_{JS}(H^{transcript}_i, H^{transcript}_j)$, whose $ij$ term equals the Jensen-Shannon distance between networks corresponding to days $i$ and $j$, on Degree phase space (figure ~\ref{fig:sim_time_matrix_Deg}) we can clearly distinguish our outlier days as particularly distant of the typical days, which are very close to each other, thus confirming our previous idea: these days we observe a collective behavior in terms of determinism. In the same figure, as also in figure ~\ref{fig:sim_time_matrix_EVC}, which shows the same for EVC phase space, we observe at least two well separated seasons, previously identified as centralized and decentralized connectedness seasons. They exhibit oscillations but are clearly distinguisable by their internal coherence (low distances detected as blue blocks on diagonal) as well as the high distances between them (yellow strips in the figures). For comparison, see \autoref{appendix: corr_matrix} for analogous figures when classical correlation coefficients are used instead of transcript synchronization.

While both phase spaces are good detecting structure in periods (highly decentralized season for instance), degree  is far better highlighting outlier days.

Of course, to choose the phase space as that of the histograms is problematic in that it adds an extra, posibly very sensitive parameter, and more generally a whole new problem: the number of bins and the binning process. Here we choose ten uniformly sized bins covering the whole range of the corresponding quantity throughout the trading year.

\begin{figure}[h!]
  \centering
    \includegraphics[width=\columnwidth]{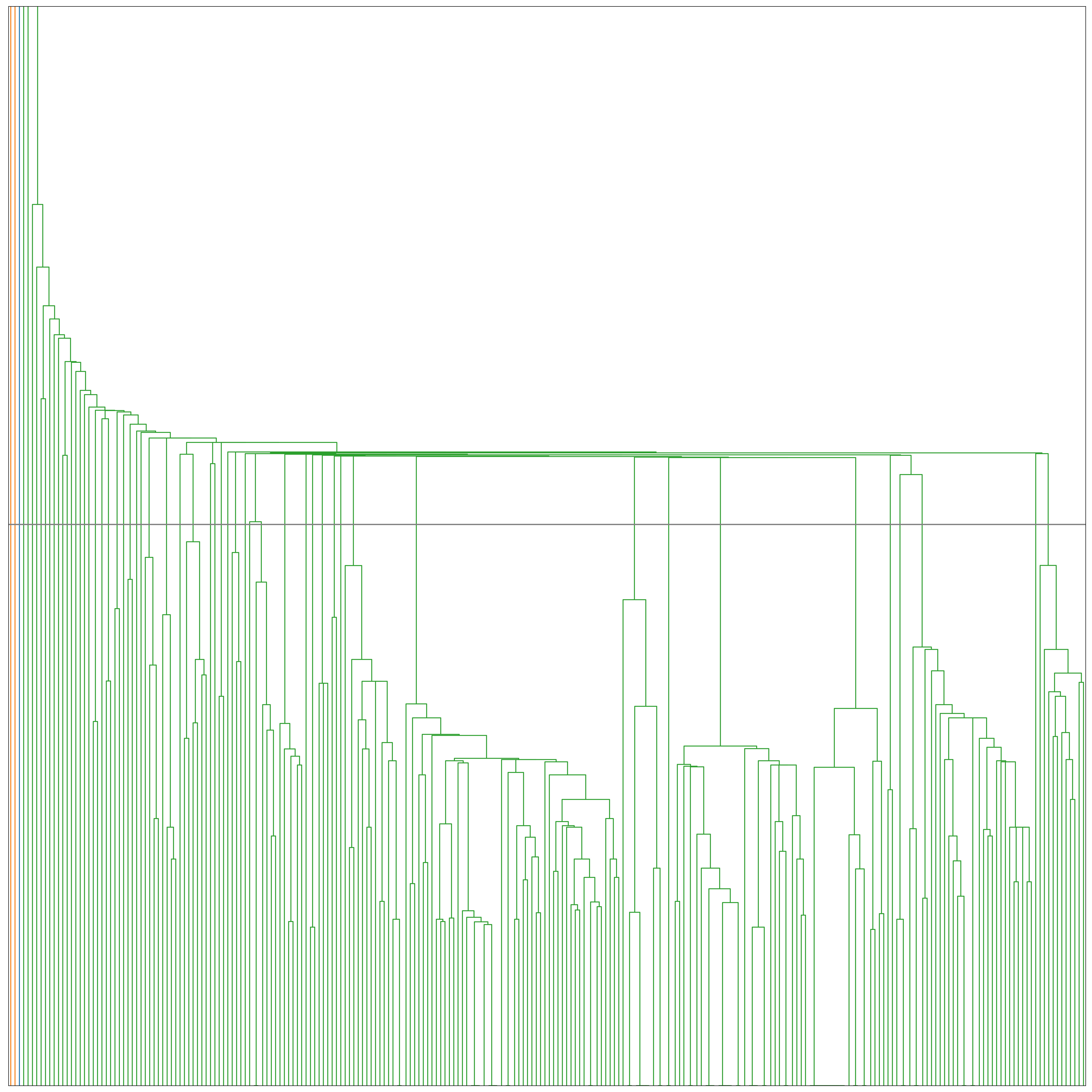}
\caption{Clustering Dendogram on Degree Phase Space with Jensen-Shannon Distance. $x$ axis represents the daily networks to be clustered, $y$ axis represents distance. Labels are omitted for better visualization. The cut-off threshold is set to 0.13 and displayed as an horizontal gray line.}\label{fig:dendogram_Deg}

\end{figure}

\begin{figure}[h!]
  \centering
    \includegraphics[width=\columnwidth]{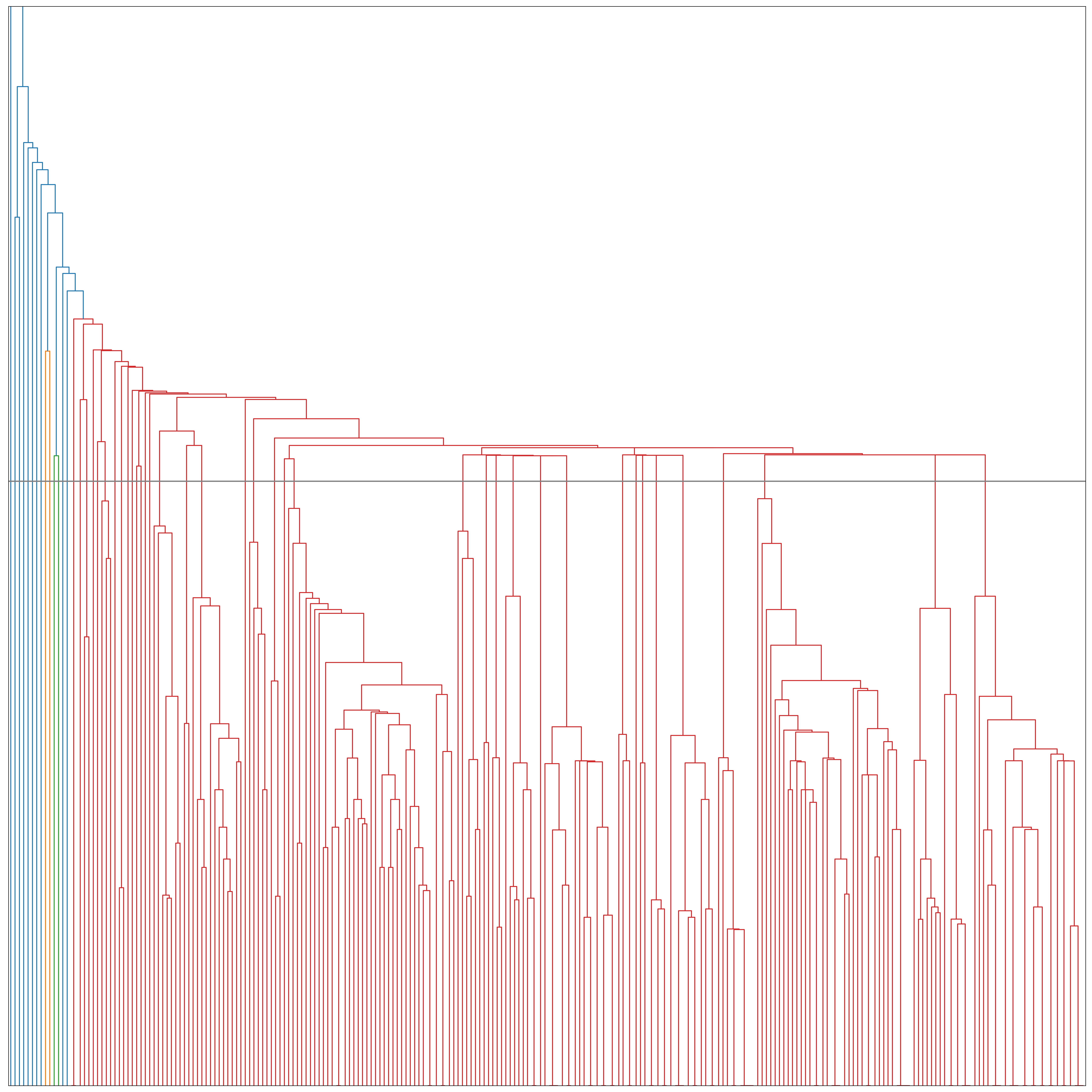}
\caption{Clustering Dendogram on EVC Phase Space with Jensen-Shannon Distance. $x$ axis represents the daily networks to be clustered, $y$ axis represents distance. Labels are omitted for better visualization. The cut-off threshold is set to 0.14 and displayed as an horizontal gray line.}\label{fig:dendogram_EVC}

\end{figure}

We can now use $\zeta$ to cluster our daily matrixes (and thus our dynamical weighted network) looking for distinctions between collective states \cite{02Marsili} \cite{12Munnix} \cite{18Pharasi}\cite{19Masuda}. For this, we use two very different algorithms: first, an aglomerative dendogram with the Jensen-Shannon distance, whose merging process is set to stop by a cut-off threshold, chosen after careful inspection of the dendograms (figures ~\ref{fig:dendogram_Deg} and ~\ref{fig:dendogram_EVC}, the threshold is shown as an horizontal gray line, and is set to 0.13 for Degree and 0.14 for EVC) and the $L^2$-based K-Means algorithm. As the dendogram gives us lots of clusters with just a couple of elements as their members, we, in order to keep the number of clusters reasonable, merge all of those with 3 elements or less into a unique set labeled as ``Noise", yet at the cost of losing some information about outliers, which as we will see will be recovered by the K-Means algorithm. The other clusters, which are our states and after the last merge into the noise set turn out to be 12 for Degree and 14 for EVC, are ordered according to the Jensen-Shannon distance between the centroid (mean) of each of them and the corresponding uniform distributions, while the clusters obtained by the K-Means algorithm are ordered by their $L^2$-norm. Thus, low states are those whose centroids lie nearer to the uniform distribution, that is, they are centralized synchronicity states, and the high states are for the same reason decentralized states. The state of a given trading day reflects then, as was our intention, the level of centralization/decentralization. The cophenetic correlation of the JS-based dendograms are $0.41$ and $0.43$ for Degree and EVC phase spaces respectively. When applying K-Means algorithm, we choose the same number of clusters as that obtained by the dendogram algorithm. 

\begin{sidewaysfigure}[h!]
    \centering
    \includegraphics[width=\textheight]{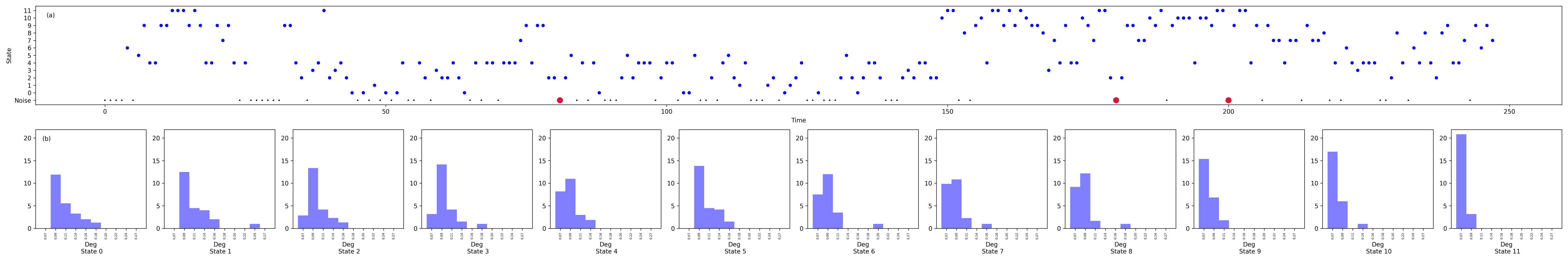}

\caption{(a) Evolution of States in Degree Phase Space and (b)$\zeta_{\JS}$-ordered Centroids, Dendogram Algorithm.}\label{fig:dendogram_States_Deg}
\bigskip
    \includegraphics[width=\textheight]{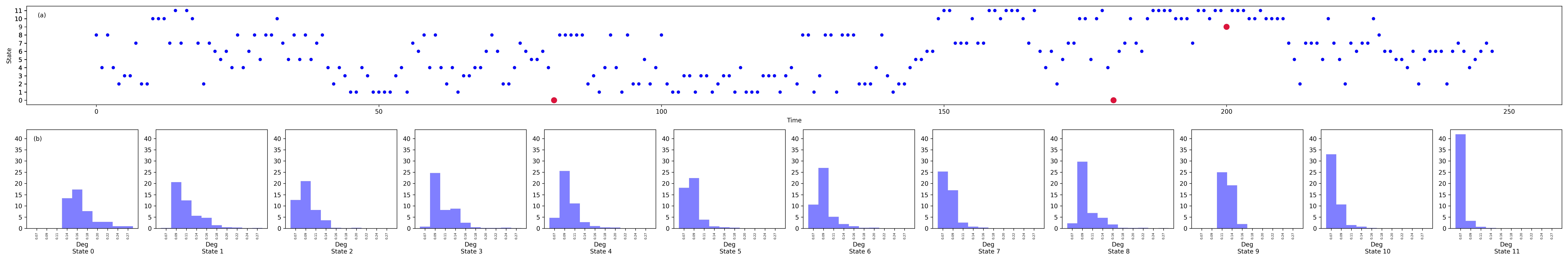}

\caption{(a) Evolution of States in Degree Phase Space and (b) $L^2$-ordered Centroids, K-Means Algorithm.}\label{fig:kmeans_States_Deg}

\end{sidewaysfigure}

\begin{sidewaysfigure}[h!]
    \centering
    \includegraphics[width=\textheight]{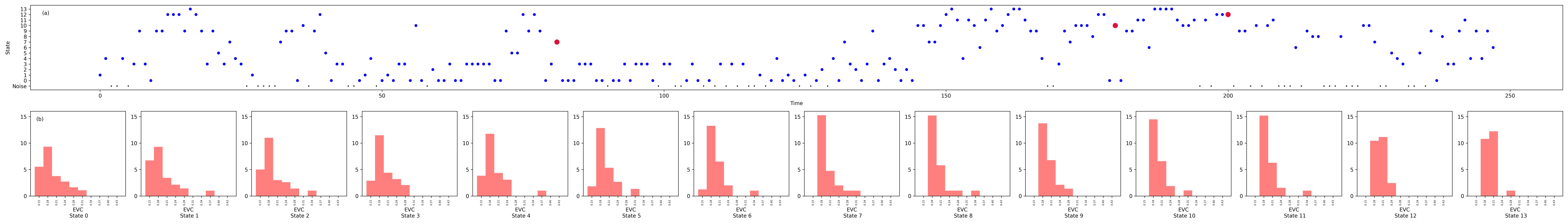}

\caption{(a) Evolution of States in EVC Phase Space and (b) $\zeta_{\JS}$-ordered Centroids, Dendogram Algorithm.}\label{fig:dendogram_States_EVC}
\bigskip
    \includegraphics[width=\textheight]{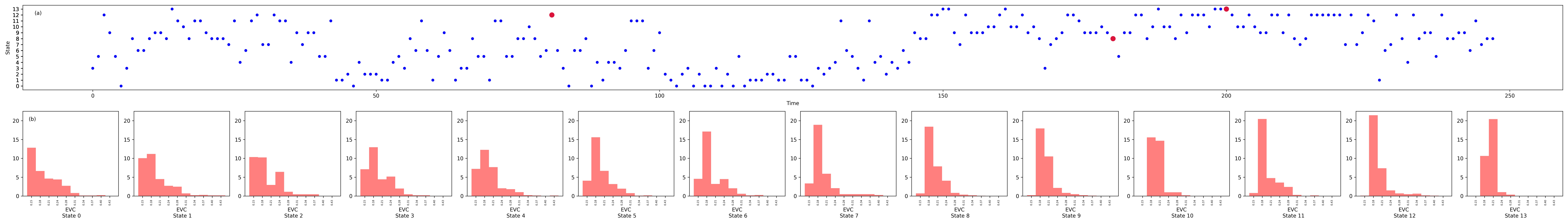}

\caption{(a) Evolution of States in EVC Phase Space and (b) $L^2$-ordered Centroids, K-Means Algorithm.}\label{fig:kmeans_States_EVC}

\end{sidewaysfigure}

In the next figures, the upper panel displays the evolution of states throughout the year: black small dots at the bottom of the graph indicate noise days when dendogram is used, blue medium-sized dots indicate typical-and-not-noisy days and red big dots indicate outlier days; when a noisy day happen to be also an outlier day, it is shown as a red big dot. The lower panel of the same figures displays the $\zeta_{\JS}$ or $L^2$-ordered centroids of the clusters (states). While both clustering methods agree in the detection of a low-states season and a high-states one, as well as in the shape of the centroids (of non-singleton clusters), they display different, complementary features. 

First of all, it is noteworthy that both phase spaces and with both clustering methods, very different in nature and using two different distances, display very similar dynamics (figures ~\ref{fig:dendogram_States_Deg}-~\ref{fig:kmeans_States_EVC}). This is a strong feature supporting our findings, as we can observe the highly decentralized season with particular persistence and clarity, as well as roughly monthly oscillations, less clear for dendogram algorithm due to the way we labeled days as noisy. 

We have now more information to further classify outlier days and grasp a little more about their nature, but first some comments are required. First of all, both clustering algorithms on both phase spaces agree in this: the last outlier day belongs to a higher state than that of the other outlier days. This allows us to conclude that the last outlier day is of a particularly decentralized nature  when compared with the others. That been said, the figures disagree in how different are those states: both algorithms finds high states for all three outlier days in EVC phase space; while this could be accepted without further thoughts for K-Means algorithm, this is unlikely for dendogram algorithm, since we would have expected the outlier days being classified as noise because of the way in which we defined the Noise set. Moreover, they are not only not classified as noise by dendogram algorithm: K-Means algorithm itself does not recognize them as outliers in any significant way, since they are not singled out as members of clusters with no more elements than themselves; and it is at this point that we should remember that EVC phase space was the one with more difficulties when it came to detect outlier days. On the contrary, in Degree phase space, which since the beginning was the strongest for outlier detection, our dendogram correctly recognizes all three outlier days as noise, which encourages us to affirm that it is to this phase space that we must turn in order to better understand outlier days. This insight is confirmed by inspecting the findings of K-Means algorithm on Degree phase space: it finds the first two oulier days belonging to a state, the lowest one, of which they are the only members, just as the last outlier day conforms a (singleton) state of its own, and a very high one indeed. Thus, outlier days, at first singled out by the individual analysis of stocks, are again and independently detected as outliers during the clustering analysis.

%If we return for a moment to figure ~\ref{fig:dynamical_deg_eigen}(a), we can see a bit more: we can confirm the decentralized nature of the last outlier day in which the corresponding row displays a particularly uniform color, and detect which stocks are those dominanting the centralized dynamics of the first two ouliers: Citigroup and Morgan Stanley from the financial sector, which strongly dominate over the whole centralized season, but also CSCO, ORCL, MSFT and F; with the excpection of F, all of them belonging to the digital technology sector.

The previous discussion should let clear that our collective analysis is able, not just to reproduce, but to further explain the nature of our individually detected outlier days as extreme manifestations of a collective behavior present throughout the trading year (centralized and decentralized synchronicity), as well as to discriminate between them in terms of that observed behavior: the first two outlier days are shown to be the most centrally synchronized of the whole year, while the last one is of a highly centralized nature, and a very peculiar one since it constitutes a market state on its own.

%In figures ~\ref{fig:dynamical_hist_deg_states_transcript} and ~\ref{fig:dynamical_hist_EVC_states_transcript} we visualize the same information in a different way, by plotting the same histograms as in figures ~\ref{fig:dynamical_hist_transcript} but this time colored according to their founded clusters-states.

%\begin{figure}[h!]
%  \centering
%    \includegraphics[width=0.48\columnwidth]{transcripts_PE_dendogram_3dhistogram_bars_States12_Deg.jpg}
%    \includegraphics[width=0.48\columnwidth]{transcripts_PE_kmeans_3dhistogram_bars_States12_Deg.jpg}

%\caption{Evolution of Histogram of Degree Colored by State. (a) Dendogram, (b) K-Means algorithm.}\label{fig:dynamical_hist_deg_states_transcript}

%\end{figure}

%\begin{figure}[h!]
%  \centering
%    \includegraphics[width=0.48\columnwidth]{transcripts_PE_dendogram_3dhistogram_bars_States14_EVC.jpg}
%    \includegraphics[width=0.48\columnwidth]{transcripts_PE_kmeans_3dhistogram_bars_States14_EVC.jpg}

%\caption{Evolution of Histogram of EVC Colored by State. (a) Dendogram, (b) K-Means algorithm.}\label{fig:dynamical_hist_EVC_states_transcript}

%\end{figure}

%%%%%%%%%%%%%%%%%%%%%%%%%%%%%%%%%%%%%%%%%%
\section{A Markov Model for State Transitions}

Finally,in order to model these state dynamics in a simple way, we briefly propose a first order Markov model for prediction of the next day state given that we know today market state. %First, let's visualize first order transition probabilities.

\begin{figure}[h!]
  \centering
  \begin{subfigure}[b]{0.40\linewidth}
    \includegraphics[width=\columnwidth]{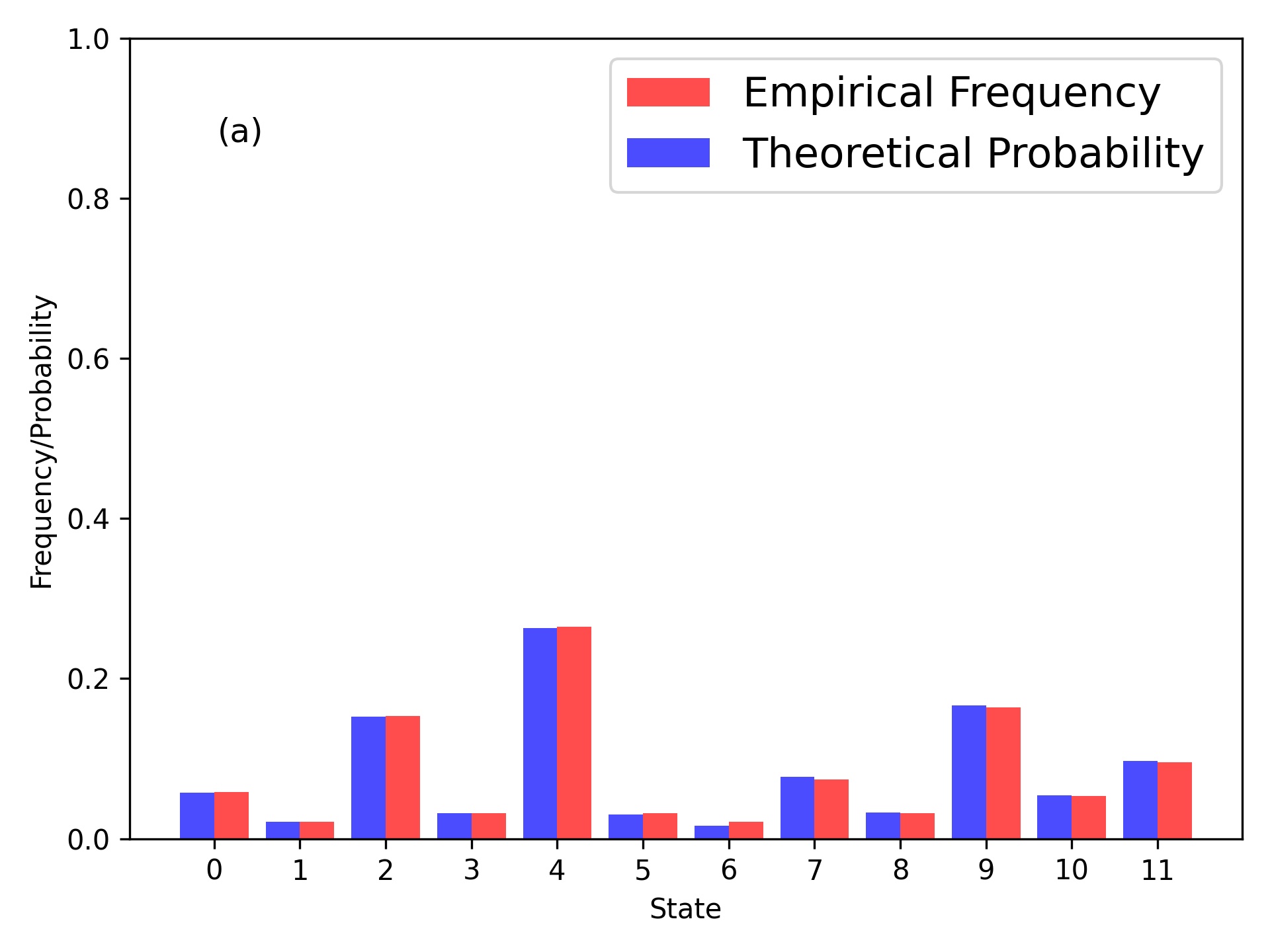}
  \end{subfigure}  
  \begin{subfigure}[b]{0.40\linewidth}
    \includegraphics[width=\columnwidth]{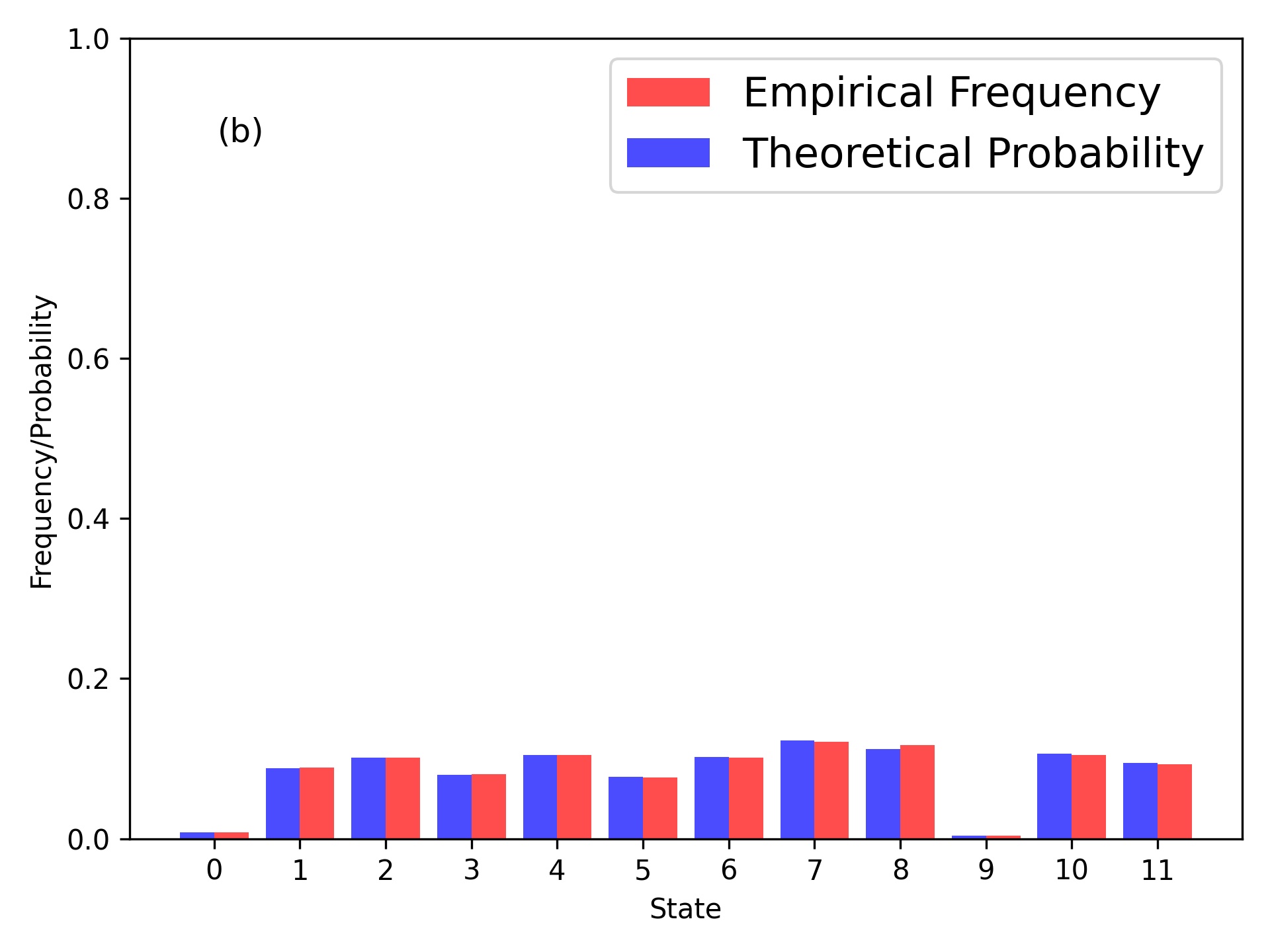}
  \end{subfigure}\hfill  
  \begin{subfigure}[b]{0.40\linewidth}
    \includegraphics[width=\columnwidth]{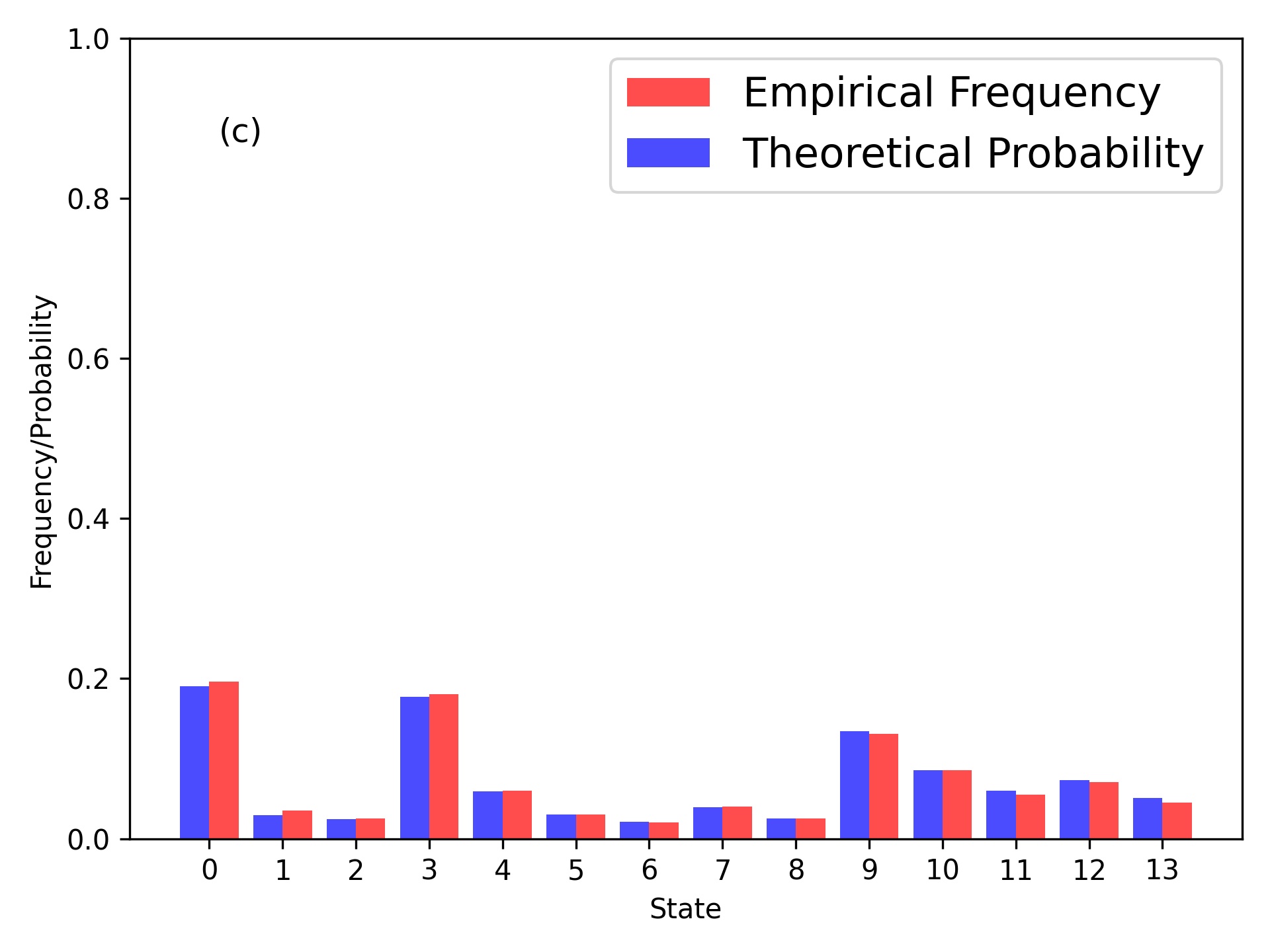}
  \end{subfigure}  
  \begin{subfigure}[b]{0.40\linewidth}
    \includegraphics[width=\columnwidth]{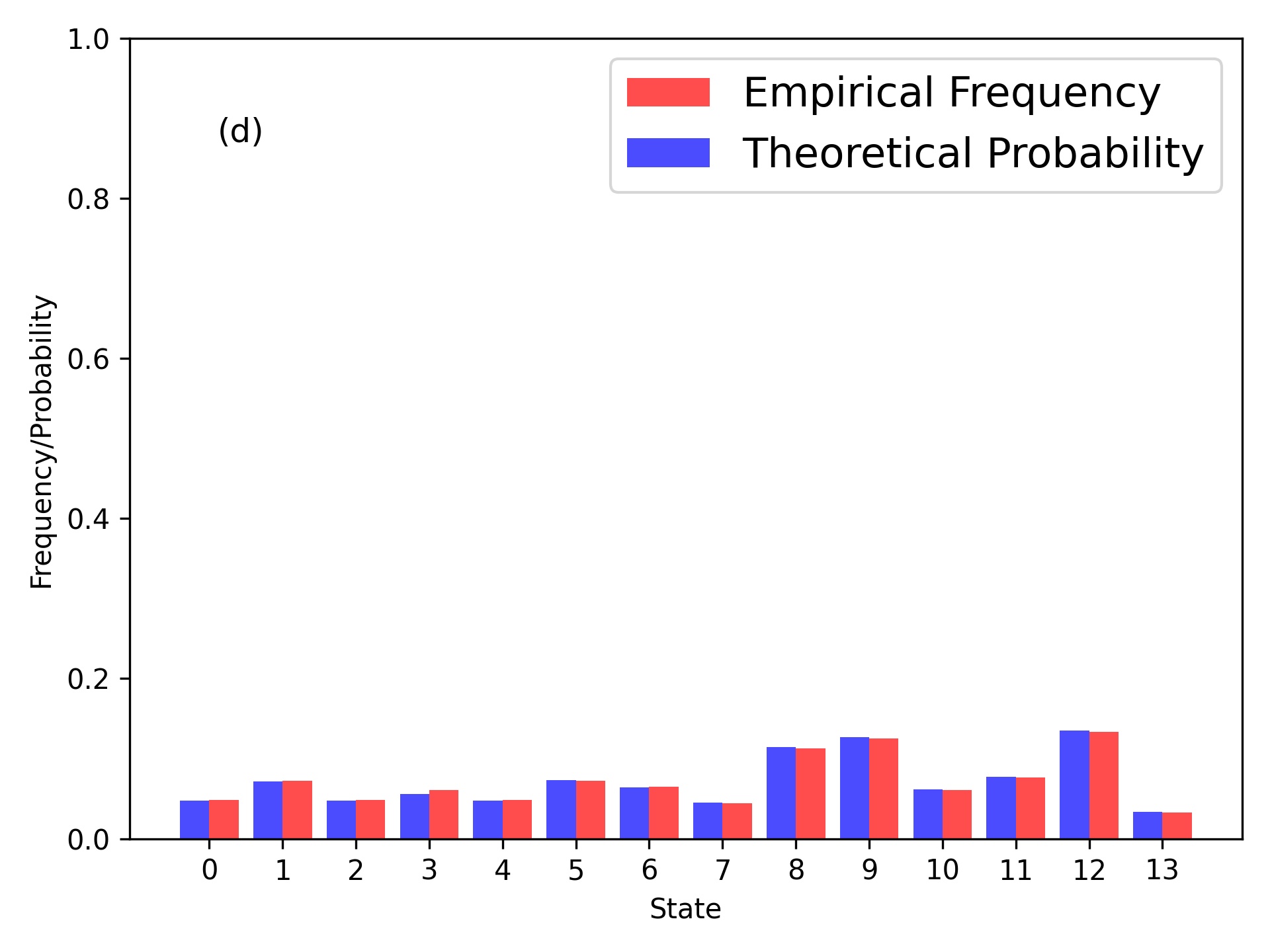}
  \end{subfigure} \hfill  
     \caption{Empirical Frequency of States and Theoretical Stationary Distribution for the Markov Model. (a) Degree, Dendogram; (b) Degree, K-Means; (c) EVC, Dendogram; (d) EVC, K-Means.}\label{fig:stationary_prob}    

\end{figure}

To check whether this Markovian approach is adequate or not we, just as in \cite{18Pharasi}, compute the empirical transition probability matrix $P$ from one state to the next, given that we know the current state, as well as the theoretical stationary probability distribution of states for such a Markov model given by the first order transition matrix $P$, that is, the probability distribution $\pi$ giving the expected probabilities of finding the market in a given state over a very long period (provided it is Markovian), and satisfying the linear equation:
\[
\pi_j =\sum_{i} P_{ij}\pi_i,
\]
and compare it to the empirical frequencies of states through the year (figure ~\ref{fig:stationary_prob}), just to find that they are indeed very similar to each other for every combination of phase spaces and clustering algorithms. We then conclude that the states dynamics is consistent with a Markov process in which enough time has passed as to reach its steady state, and that to guess tomorrow state given knowledge about today state is in general a reasonable bet.   

A considerable part of our work on ordinal patterns has been done with the Python package ordpy \cite{21Pessa}, while clustering analysis has been carried out through scikit-learn \cite{11Pedregosa}. We also have made extensive, though elementary, use of NumPy \cite{20Harris}.

%%%%%%%%%%%%%%%%%%%%%%%%%%%%%%%%%%%%%%%%%%
\section{Conclusions}

In order to analyze collective states dynamics of stocks in high-frequency digital markets, and to overcome the limitations of correlation matrixes in detecting non-linear interactions in noisy time series, we proposed to study transcripts synchronization dynamical networks and their eigenvector centrality and degree vectors distributions. 

After measuring different information theoretic quantities on the daily ordinal pattern series of individual stocks and detecting three outlier, semi-deterministic days in most of them and discrete levels in some, we have shown them to be extreme manifestations of a collective, emergent behavior not entirely reflected in individual dynamics. Applying two very different clustering algorithms, we were able to detect specific, persistent and quantitatively distinguishable market states throughout the one-year period of study, as well as two well defined and quantitatively distinguishable seasons of the trading year, characterized by their degree of centralized/decentralized synchronicity, with remarkable similar results for both algorithms. We also succesfully classified our previously found outlier days in terms of centralized and decentralized synchronicity. Finally, we showed that state transitions dynamics can be well described as a simple first order Markov process.

Of course, our work has several limitations that ought to be highlighted. As already mentioned, to choose the phase space as that of the histograms of EVC and Degree leaves open the question of the correct number of bins to be used. Also, it would be desirable to find a more objective criterion to choose the number of clusters; various purely quantitative ones are discussed in the litearature (see for a summary \cite{13Arbelaitz}), but none of those are convincent from our viewpoint; and ultimately, as explained in \cite{11Luxburg}, clustering analysis is a problem dependent process and should not be subordinated to an abstract, global score. As our aim in this work was to propose and illustrate a methodology, as well as to adapt the network-clustering pipeline often mentioned thorughout this work \cite{19Masuda} to high-frequency digital markets, we did not deepen into this questions; instead, in both cases we contented ourselves with confirming the robustness of our results by varying the number of bins and dendogram threshold and founding similar qualitative results in a reasonable range of values and with two very different clustering algorithms for our particular data set. 

It would be desirible to have an ecomomic explanation for the behavior here observed; unfourtunately, that is significantly difficult for high-frequency digital markets, because algorithms are particularly opaque in their trading decisions \cite{14Lewis} \cite{15Pasquale}. At the moment, we can just stand for a phenomenological approach such as that of our paper.

\begin{appendices}

\section{Correlation Matrixes}\label{appendix: corr_matrix}

For the sake of comparison, and to make clear why we use transcript synchronicity as our pairwise coupling measure instead of the more classic and straightforward correlation coefficient, we include here a few figures similar to those displayed above, but this time for correlation matrixes. However, since we are talking here about synchronization, we use the absolute value of such correlation coefficients, since a correlation coefficient equal to $-1$ should be understood as two perfectly (linearly) synchronized time series.

\begin{figure}[h!]
  \centering
    \includegraphics[width=0.48\columnwidth]{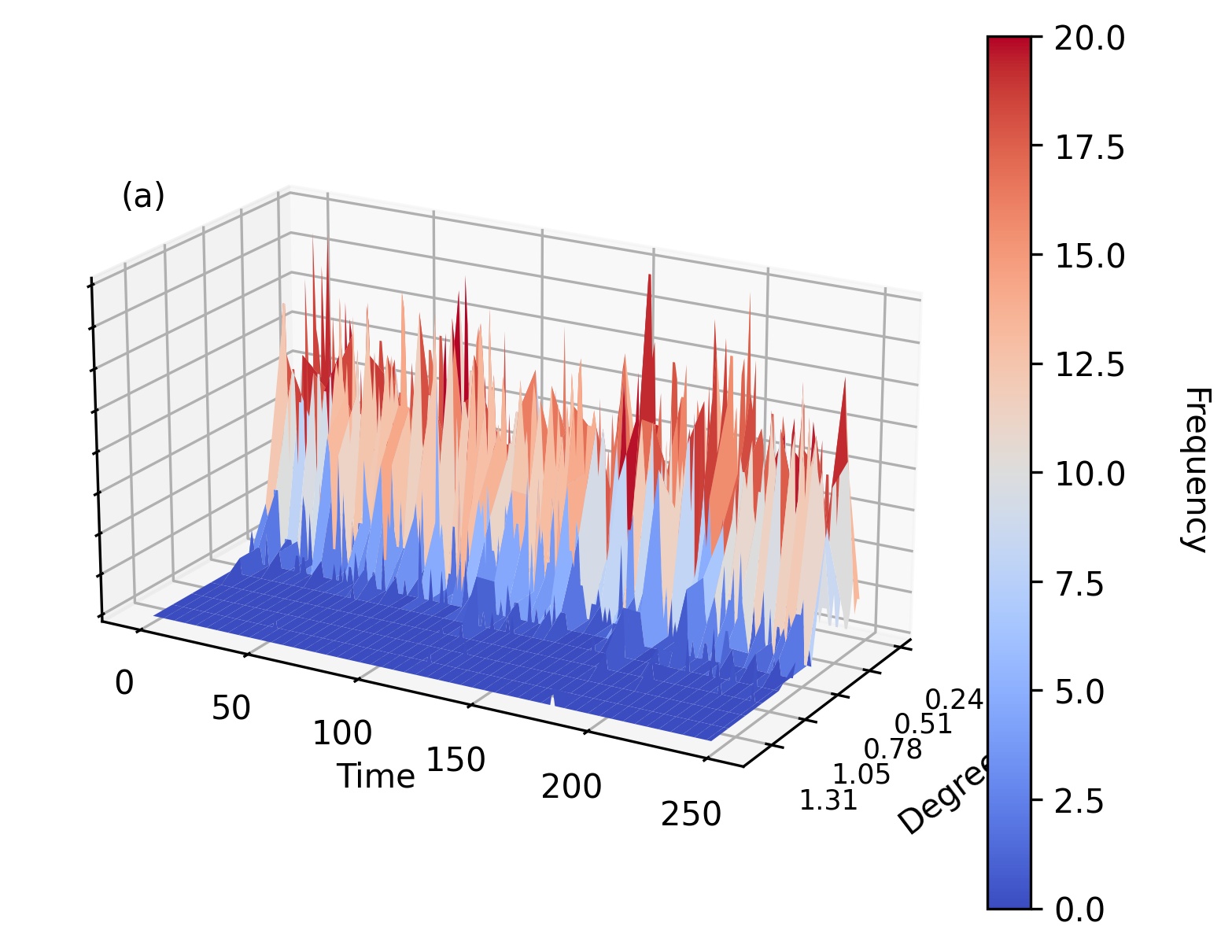}
    \includegraphics[width=0.48\columnwidth]{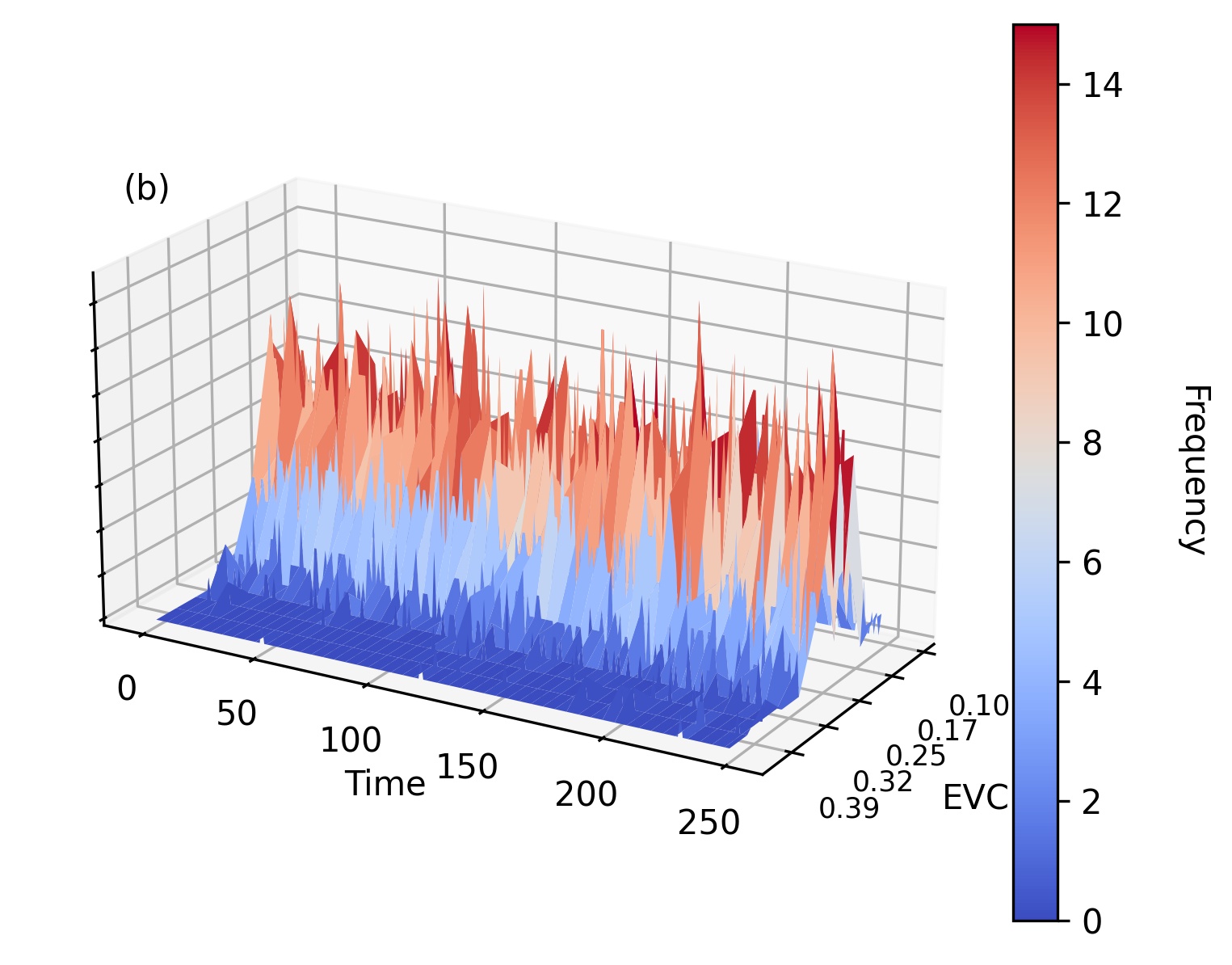}    
\caption{Evolution Through the Trading Year of Histograms of (a) Degree and (b) Eigenvector Centrality for Correlation Dynamical Network.}\label{fig:dynamical_hist_corr}
\end{figure}

In figure ~\ref{fig:dynamical_hist_corr} we can see that, althought some structure is still present in the evolution of the histogram of Degree for correlation matrixes, its presence is less clear than that observed in ~\ref{fig:dynamical_hist_transcript}, while the histogram of EVC is not useful at all when correlation coefficients are used.

\begin{figure}[h!]
  \centering
    \includegraphics[width=0.48\columnwidth]{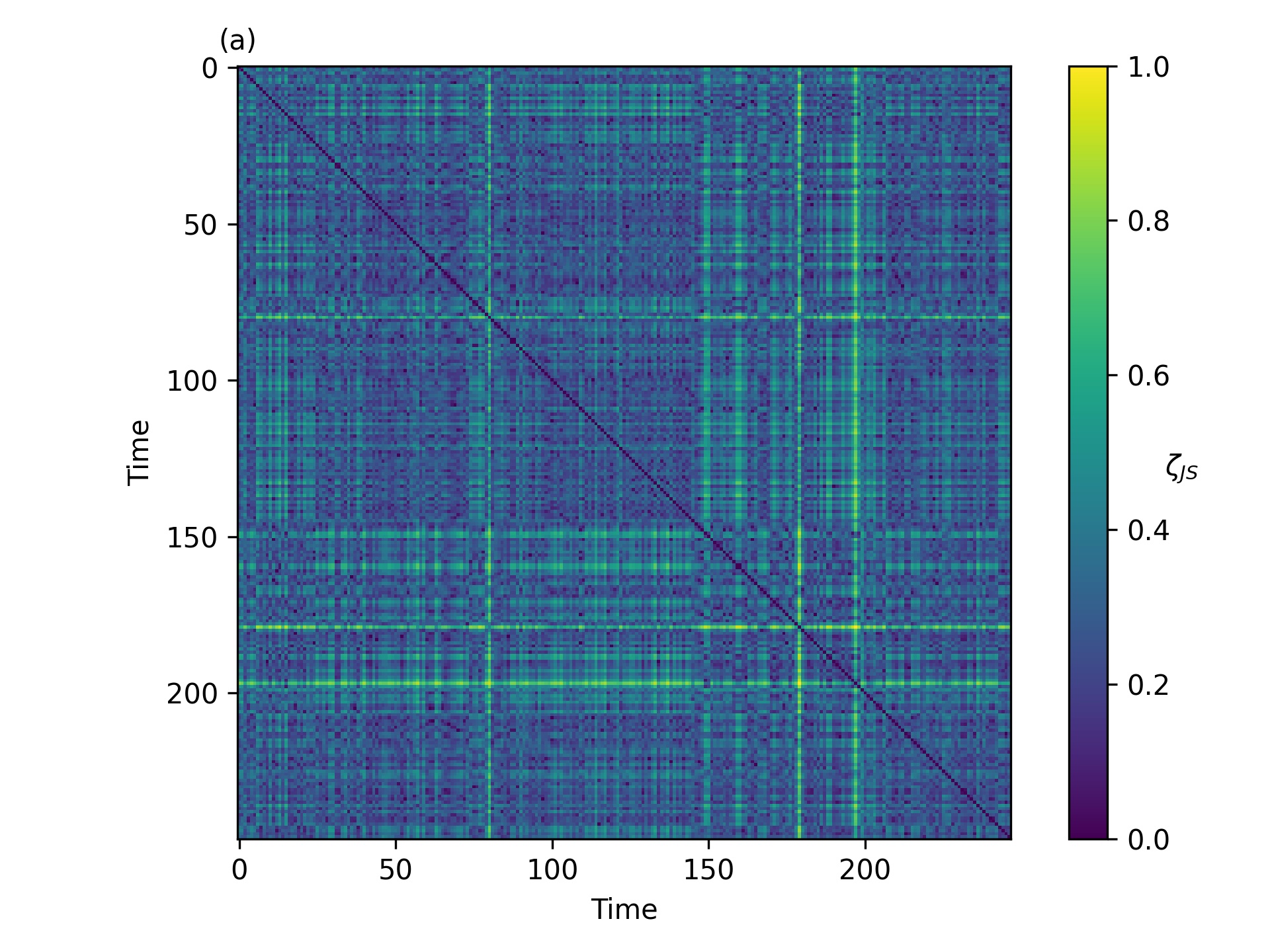}
    \includegraphics[width=0.48\columnwidth]{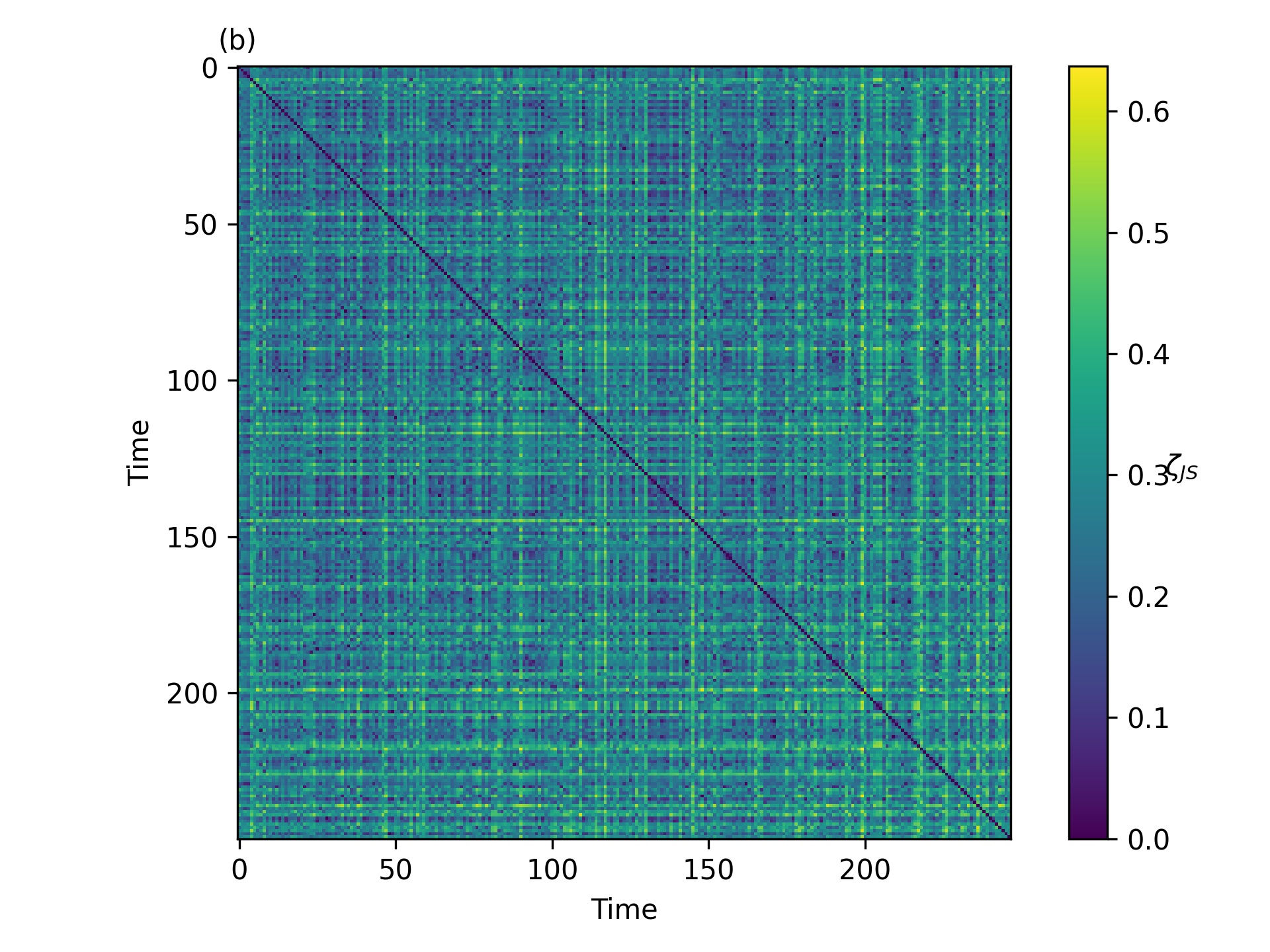}
\caption{Jensen-Shannon Distance Matrix in (a) Degree and (b) EVC Phase Spaces. Its $ij$ term equals the Jensen-Shannon distance between correlation networks corresponding to days $i$ and $j$.}\label{fig:sim_time_matrix_corr}
\end{figure}

This should be clearer in figure ~\ref{fig:sim_time_matrix_corr}, which shows the matrix of Jensen-Shannon distances between daily correlation matrixes, analogously to figures ~\ref{fig:sim_time_matrix_Deg} and ~\ref{fig:sim_time_matrix_EVC}. The first panel of this figure still correctly detects oulier days clearly enough, but the knowledge of the centralized and decentralized seasons is almost totally lost, while the second panel is too noisy to conclude anything. Consequently, clustering analysis yields poorer results when compared to our previous transcript-based analysis.

\section{Multiscale Analysis}\label{appendix: multiscale}

In \autoref{sec: theory} we made the statement that our results are robust to variations of the time lag parameter $l$. As already mentioned, in \cite{20Olivares_1} the authors make it clear that a multiscale analysis is unavoidable if we want to guarantee the validity of our results. Thus, we plot here a couple of figures obtained when $l = 5, 25, 50, 75$ and $100$ and everything else is kept as before. For brevity and space, we limit ourselves to plot the distance matrixes analogous to figures ~\ref{fig:sim_time_matrix_Deg} and ~\ref{fig:sim_time_matrix_EVC}. 

\begin{figure}[h!]
  \centering
  \includegraphics[width=0.19\columnwidth]{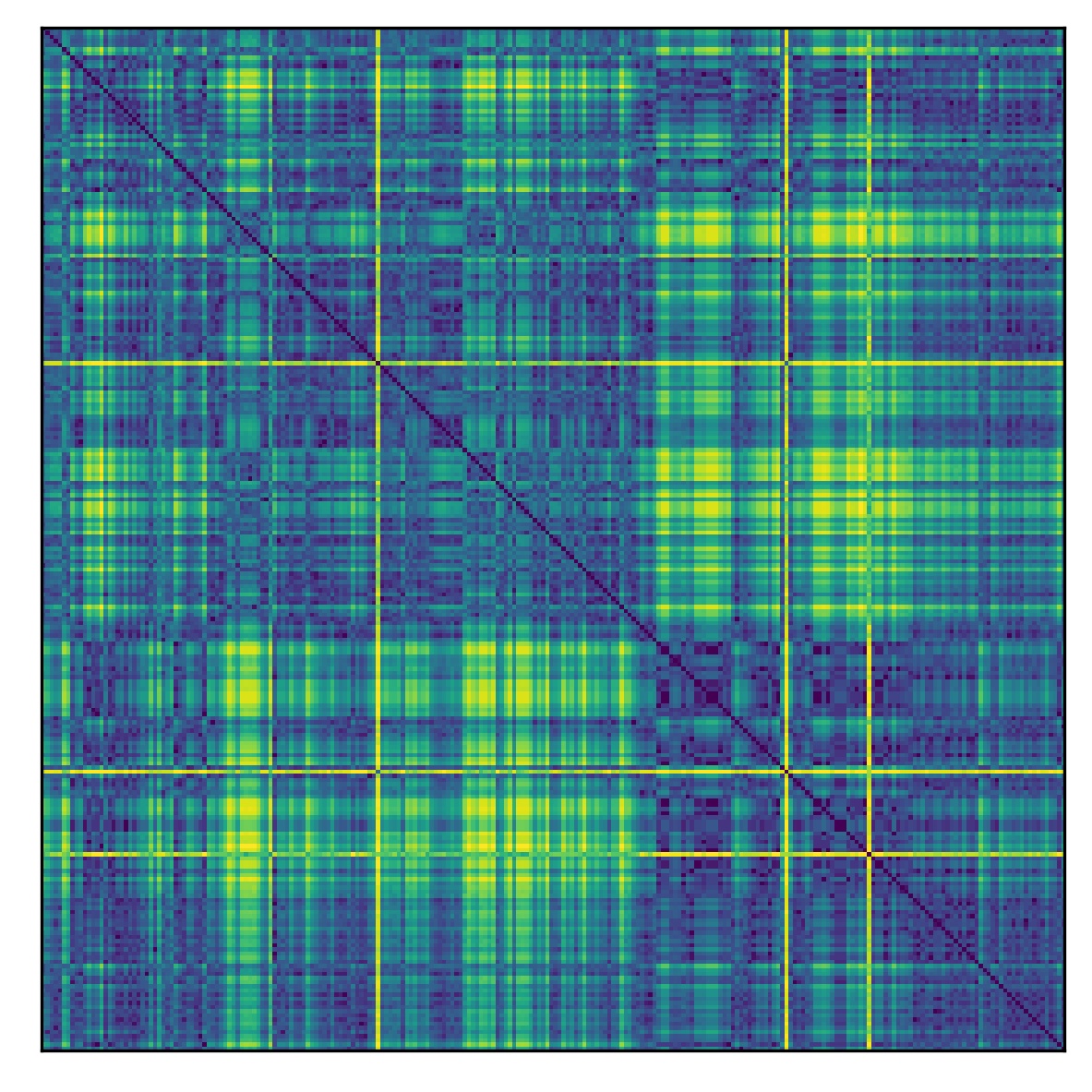}
  \includegraphics[width=0.19\columnwidth]{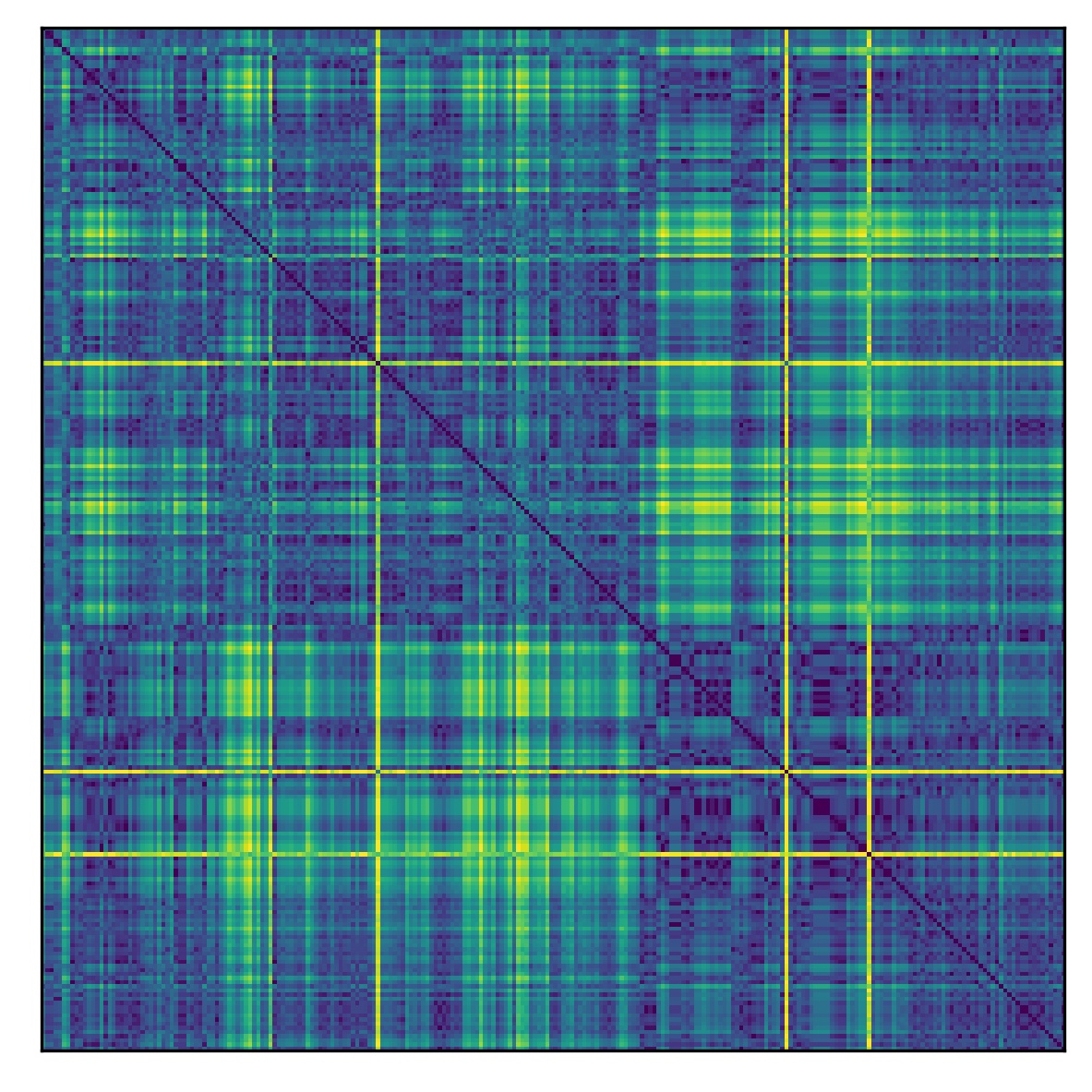}
  \includegraphics[width=0.19\columnwidth]{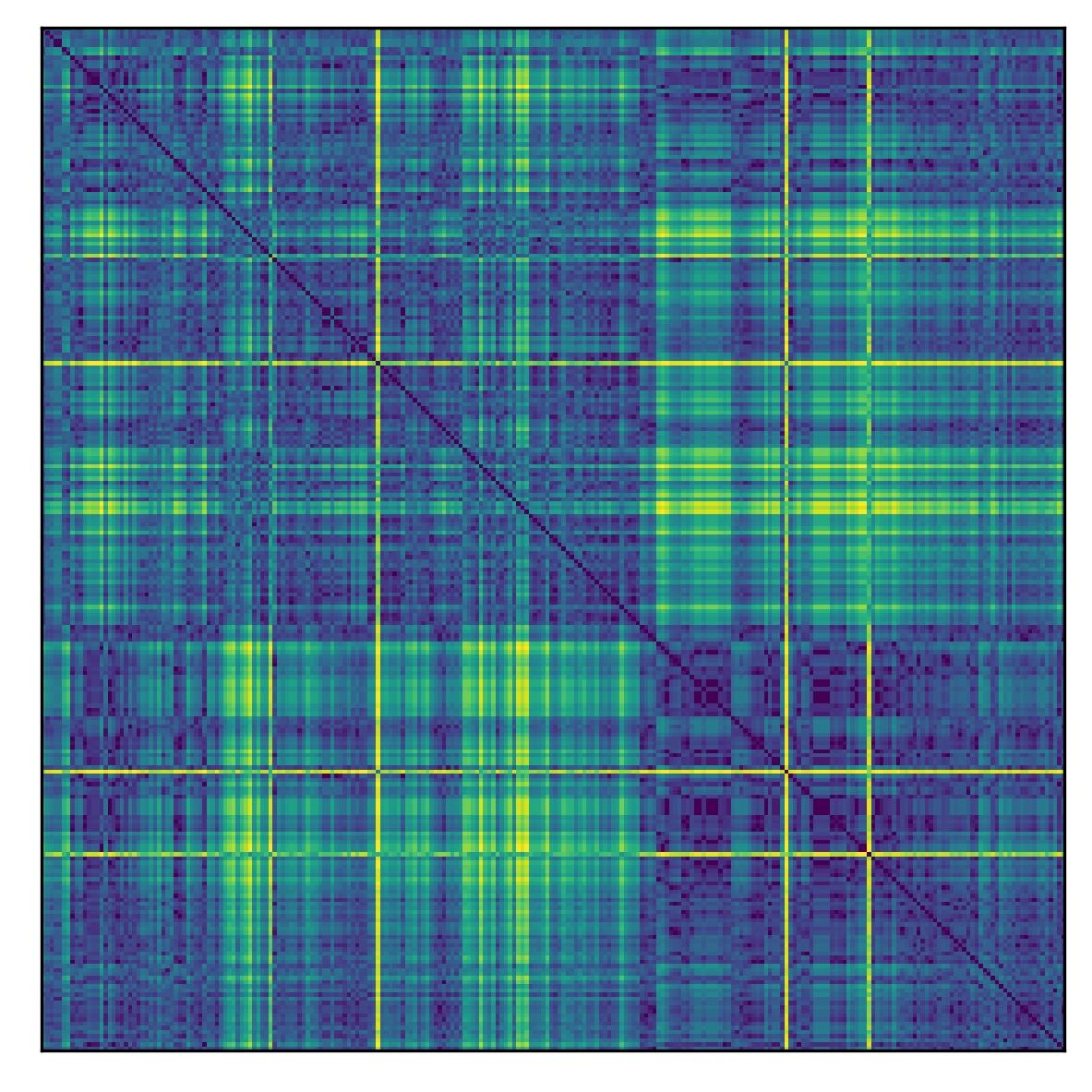}
  \includegraphics[width=0.19\columnwidth]{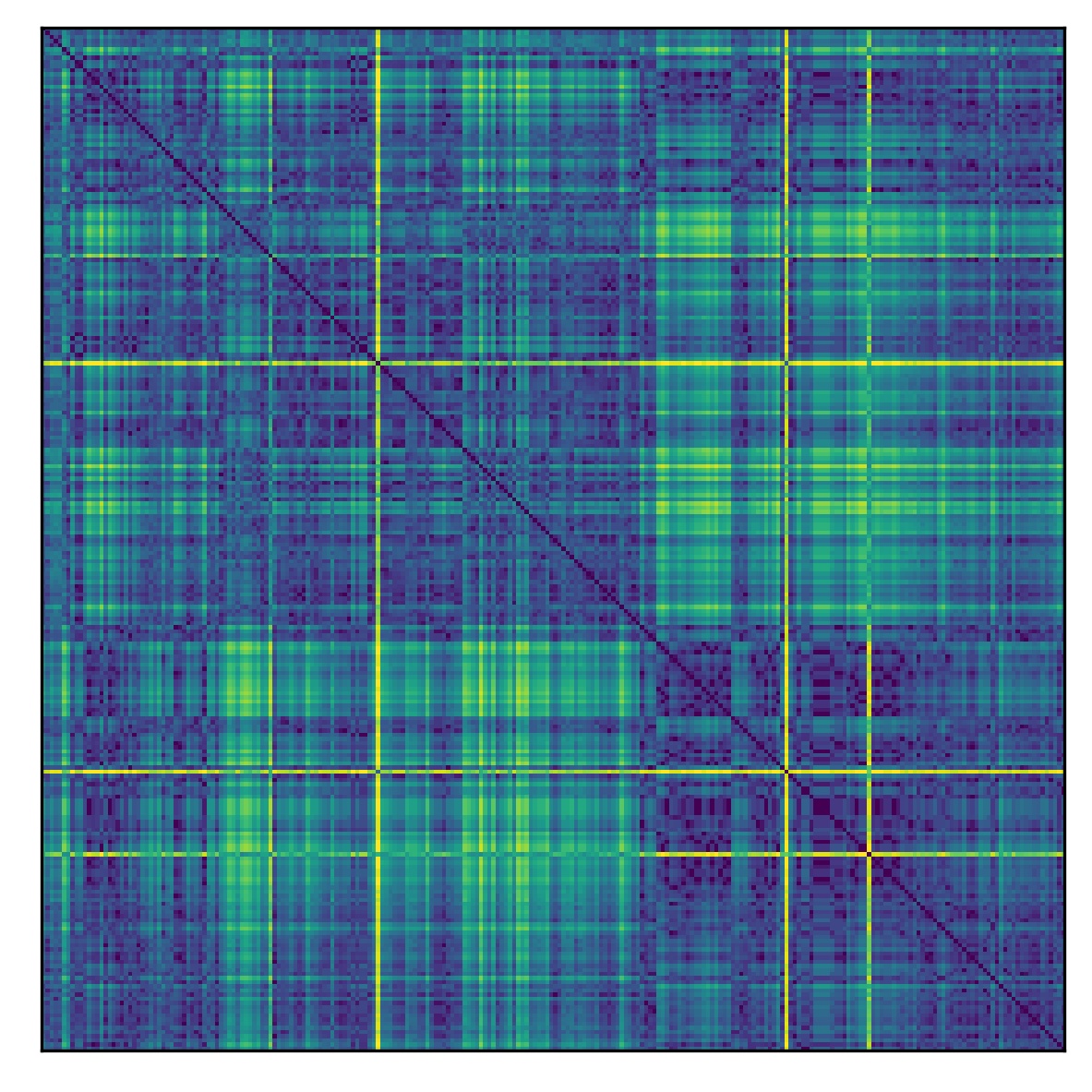}
  \includegraphics[width=0.19\columnwidth]{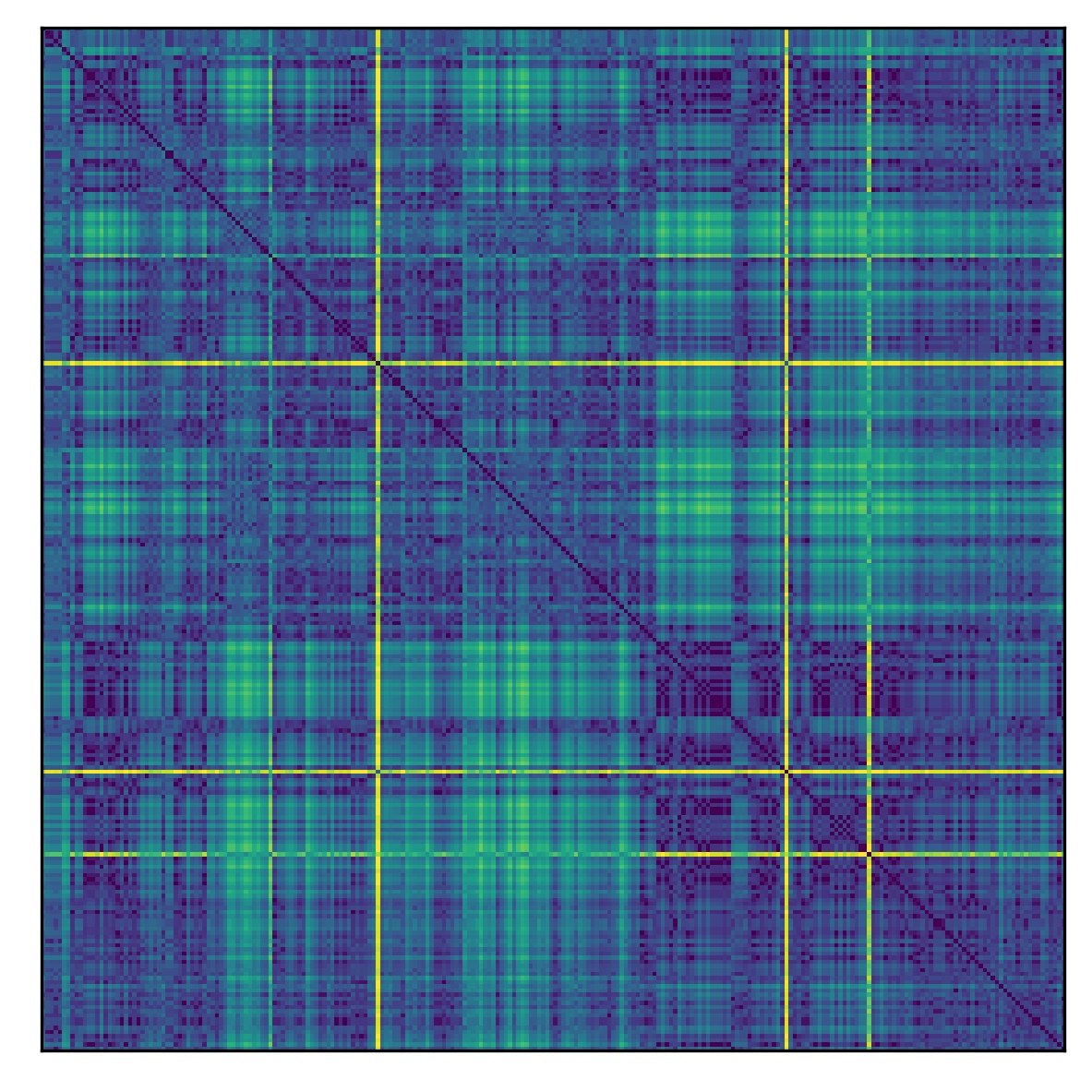}
  \includegraphics[width=0.19\columnwidth]{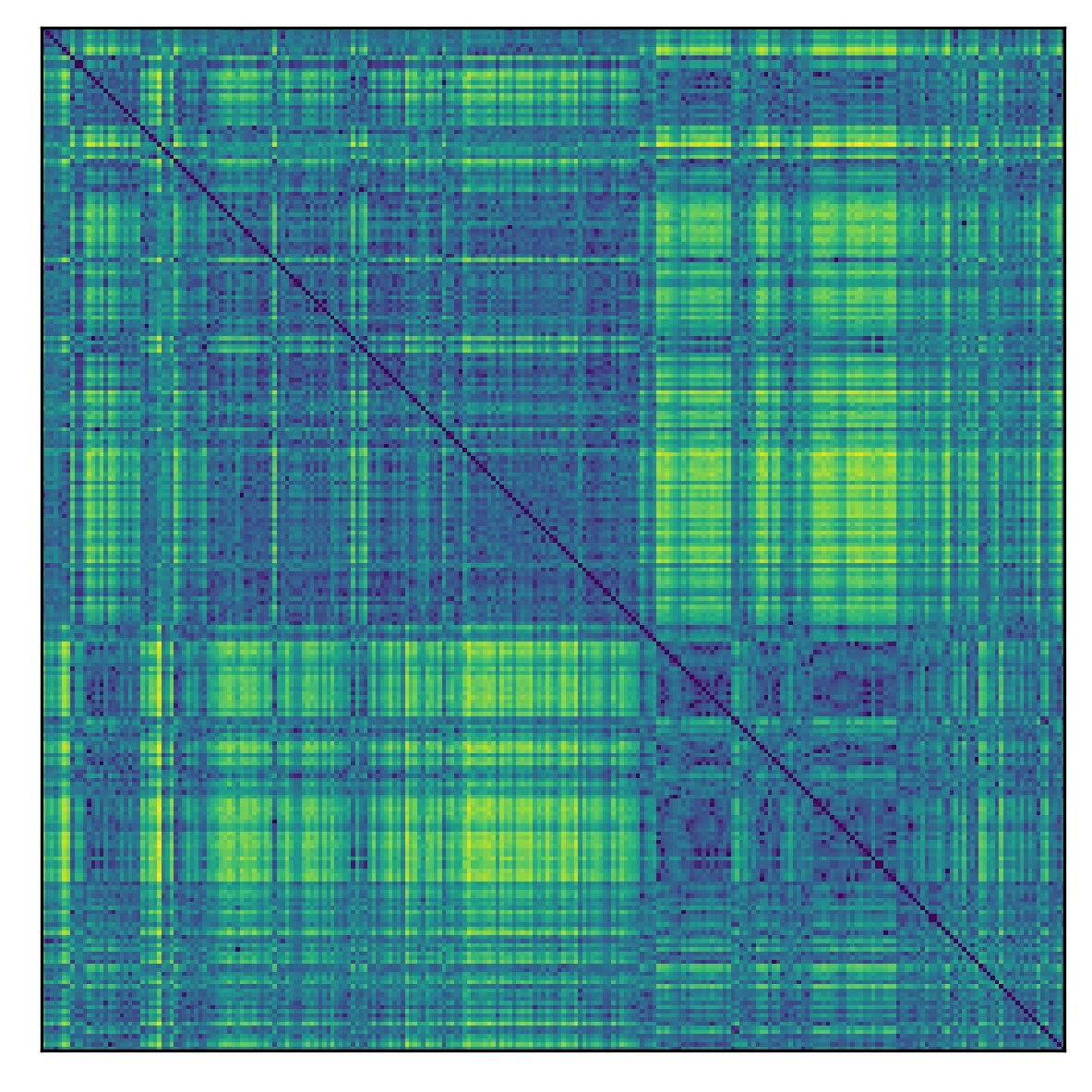}
  \includegraphics[width=0.19\columnwidth]{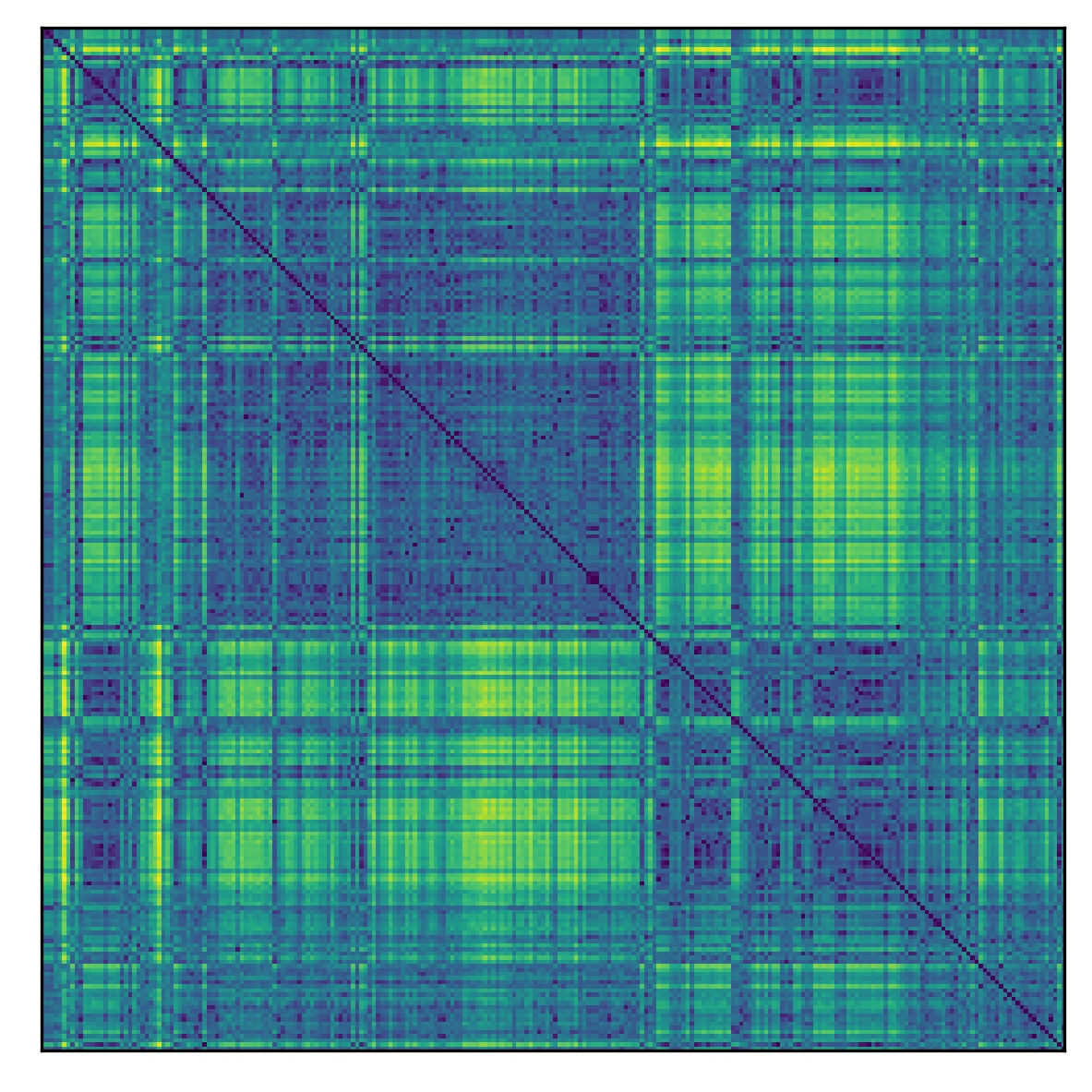}
  \includegraphics[width=0.19\columnwidth]{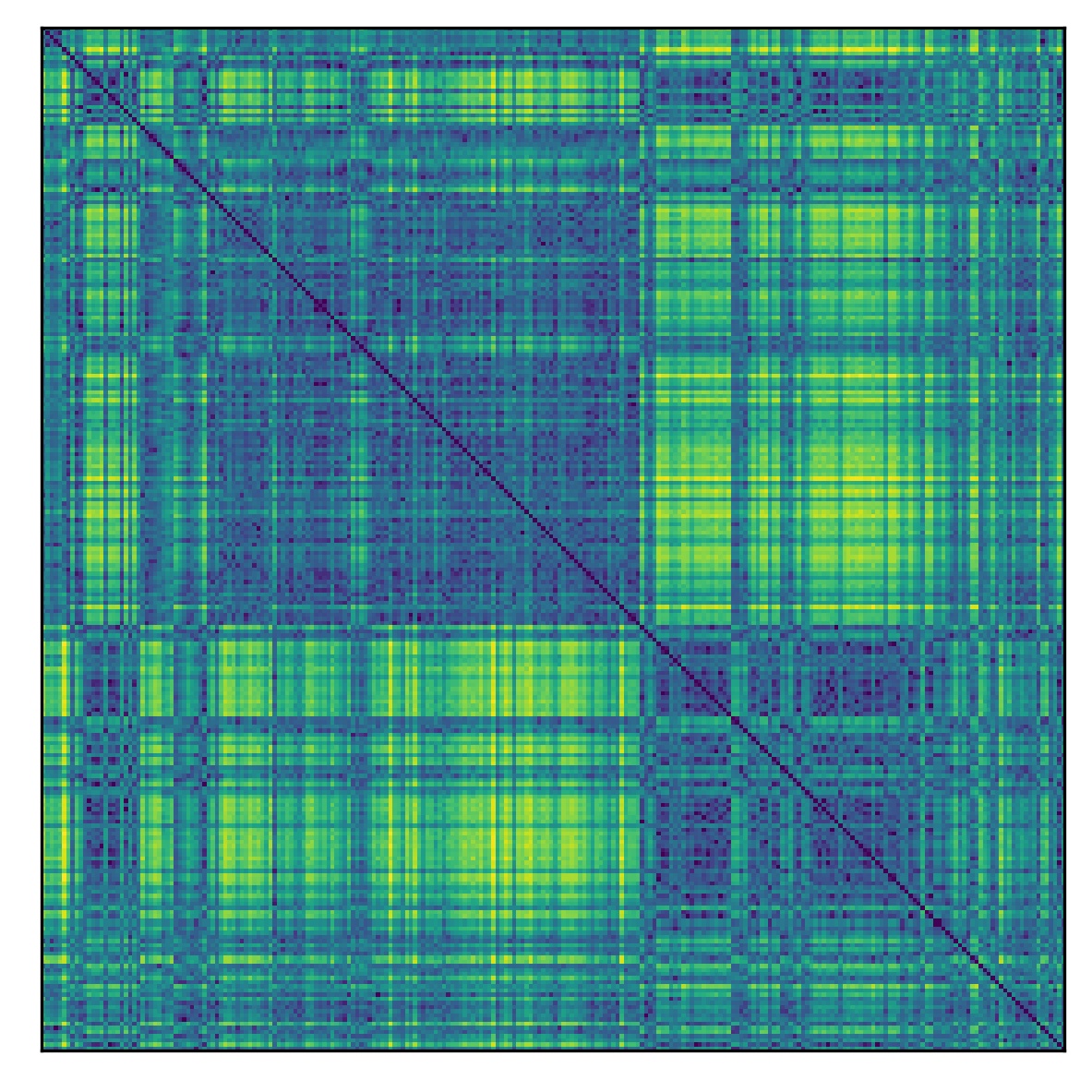}
  \includegraphics[width=0.19\columnwidth]{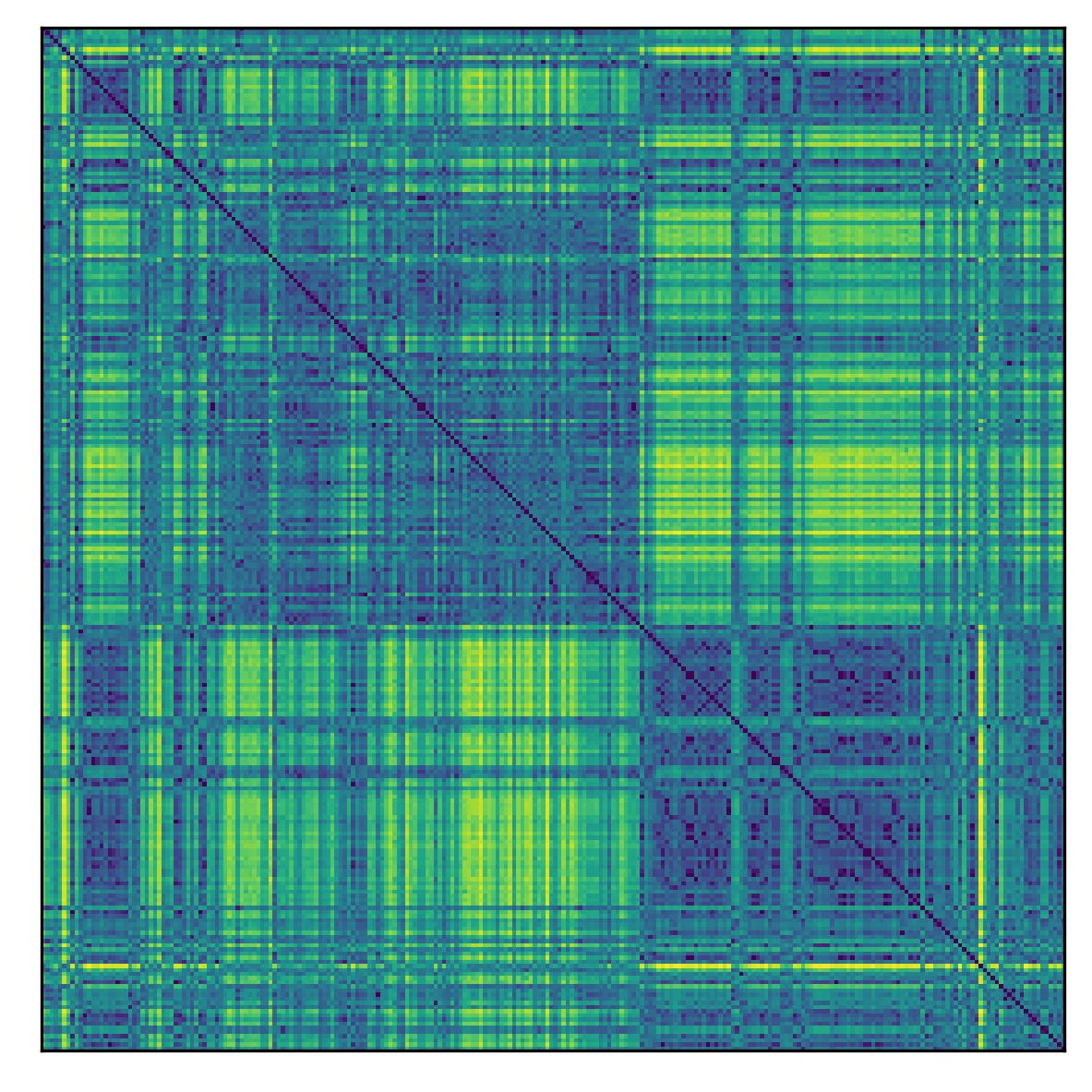}
  \includegraphics[width=0.19\columnwidth]{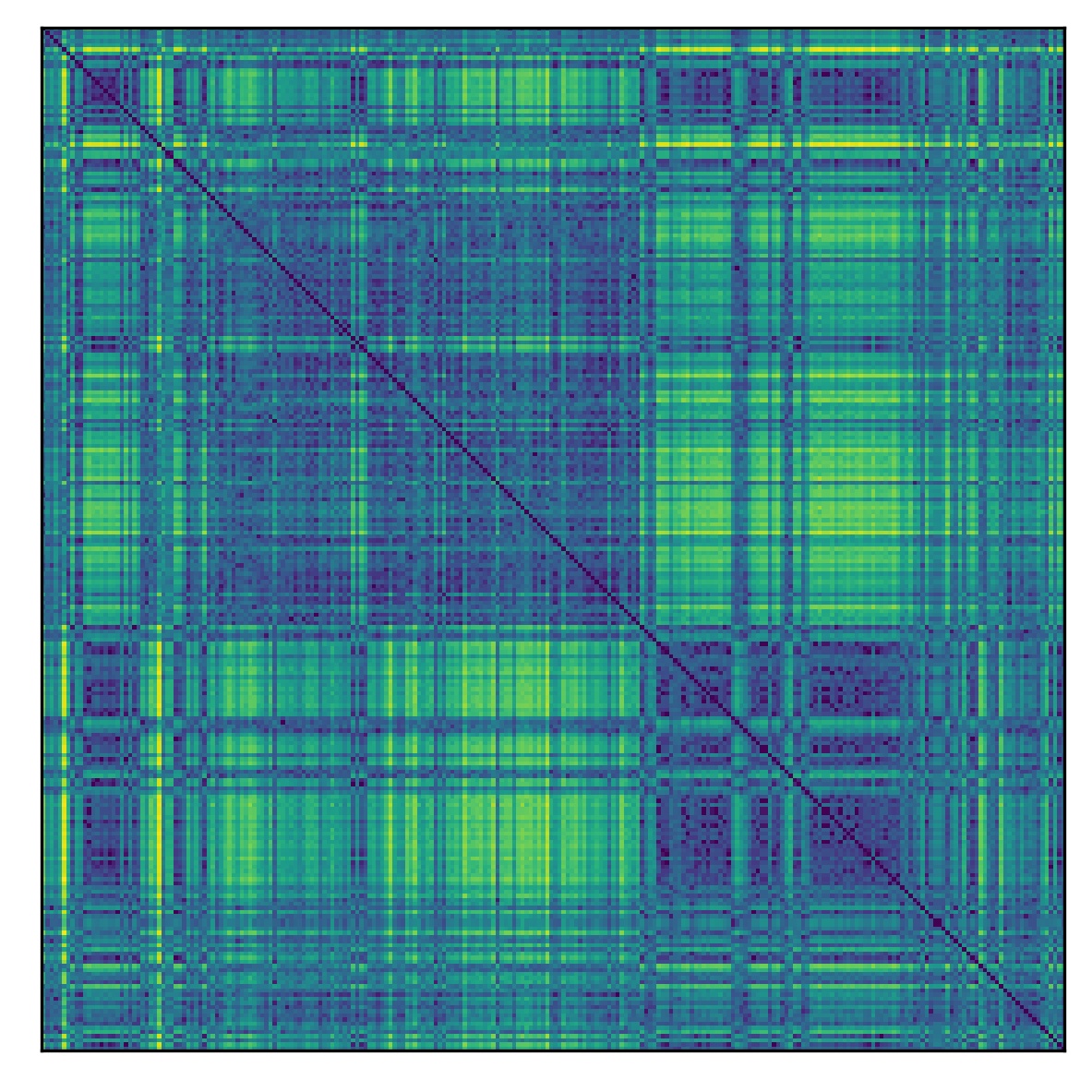}
    
\caption{Jensen-Shannon Distance Matrixes in Degree (upper panel) and EVC (lower panel) Phase Spaces. Its $ij$ term equals the Jensen-Shannon distance between transcript sychronization networks corresponding to days $i$ and $j$ for $l = 5, 25, 50, 75$ and $100$, plotted in that order from left to right.}\label{fig:sim_time_matrix_multiscale}
\end{figure}

Figure ~\ref{fig:sim_time_matrix_multiscale}, displays Jensen-Shannon distance matrixes for Degree (upper panel) and EVC (lower panel) phase spaces. It is clear that these matrixes are pretty similar to those just refered, detecting outlier days as well as highly centralized and decentralized seasons. Labels and colorbars have been removed to improve visualization. Clustering analysis yields similar results.

\end{appendices}

%%%%%%%%%%%%%%%%%%%%%%%%%%%%%%%%%%%%%%%%%%
\vspace{6pt} 

\subsection*{Author Contributions}
Conceptualization, M.L.; methodology, M.L. and R.M.; software, M.L.; formal analysis, M.L.; investigation, M.L.; resources, R.M.; data curation, M.L. and R.M.; writing---original draft preparation, M.L.; writing---review and editing, M.L. and R.M.; visualization, M.L.; supervision, M.L. and R.M. 

\subsection*{Funding}
This research received no external funding

\subsection*{Data Availability}
The data presented in this study are available \cite{21data}.

\subsection*{Conflicts of Interest}
The authors declare no conflict of interest.

\bibliographystyle{unsrt}
\bibliography{ord_sync_bib}

\begin{thebibliography}{10}

\bibitem{99Mantegna}
Rosario~N. Mantegna.
\newblock Hierarchical structure in financial markets.
\newblock {\em The European Physical Journal B}, 11:193--197, 1999.

\bibitem{03Onnela}
J.-P. Onnela, A.~Chakraborti, K.~Kaski, and J.~Kert\'esz.
\newblock Dynamic asset trees and black monday.
\newblock {\em Physica A: Statistical Mechanics and its Applications},
  324(1):247--252, 2003.
\newblock Proceedings of the International Econophysics Conference.

\bibitem{02Marsili}
Matteo Marsili.
\newblock Dissecting financial markets: sectors and states.
\newblock {\em Quantitative Finance}, 2(4):297--302, 2002.

\bibitem{12Munnix}
M.~C. M\"{u}nnix, T.~Shimada, R.~Sch\"afer, F.~Leyvraz, T.~H. Seligman,
  T.~Guhr, and H.~E. Stanley.
\newblock Identifying states of a financial market.
\newblock {\em Scientific Reports}, 2, 2012.

\bibitem{15Chetalova}
Desislava Chetalova, Rudi Sch\"{a}fer, and Thomas Guhr.
\newblock Zooming into market states.
\newblock {\em Journal of Statistical Mechanics: Theory and Experiment},
  2015(1):P01029, jan 2015.

\bibitem{19Masuda}
H.~Masuda and P.~Holme.
\newblock Detecting sequences of system states in temporal networks.
\newblock {\em Scientific Reports}, 9:795, 2019.

\bibitem{21Chakraborti}
Anirban Chakraborti, Hrishidev, Kiran Sharma, and Hirdesh~K Pharasi.
\newblock Phase separation and scaling in correlation structures of financial
  markets.
\newblock {\em Journal of Physics: Complexity}, 2(1):015002, nov 2020.

\bibitem{20Chakraborti}
Anirban Chakraborti, Kiran Sharma, Hirdesh~K Pharasi, K~Shuvo Bakar, Sourish
  Das, and Thomas~H Seligman.
\newblock Emerging spectra characterization of catastrophic instabilities in
  complex systems.
\newblock {\em New Journal of Physics}, 22(6):063043, jun 2020.

\bibitem{18Pharasi}
Hirdesh~K. Pharasi, Kiran Sharma, Rakesh Chatterjee, Anirban Chakraborti,
  Francois Leyvraz, and Thomas~H. Seligman.
\newblock Identifying long-term precursors of financial market crashes using
  correlation patterns.
\newblock {\em New Journal of Physics}, 20(10):103041, nov 2018.

\bibitem{21Musciotto}
Federico Musciotto, Jyrki Piilo, and Rosario~N. Mantegna.
\newblock High-frequency trading and networked markets.
\newblock {\em Proceedings of the National Academy of Sciences}, 118(26), 2021.

\bibitem{17Kim}
M.~Kim and H.~Sayama.
\newblock Predicting stock market movements using network science: an
  information theoretic approach.
\newblock {\em Applied Network Science}, 2, 2017.

\bibitem{12Zanin}
Massimiliano Zanin, Luciano Zunino, Osvaldo~A. Rosso, and David Papo.
\newblock Permutation entropy and its main biomedical and econophysics
  applications: A review.
\newblock {\em Entropy}, 14(8):1553--1577, 2012.

\bibitem{02Bandt}
Christoph Bandt and Bernd Pompe.
\newblock Permutation entropy: A natural complexity measure for time series.
\newblock {\em Phys. Rev. Lett.}, 88:174102, Apr 2002.

\bibitem{12Cysarz}
D.~Cysarz, P.~{Van Leeuwen}, F.~Edelh\"{a}user, N.~Montano, and A.~Porta.
\newblock Binary symbolic dynamics classifies heart rate variability patterns
  linked to autonomic modulations.
\newblock {\em Computers in Biology and Medicine}, 42(3):313--318, 2012.

\bibitem{12Parlitz}
U.~Parlitz, S.~Berg, S.~Luther, A.~Schirdewan, J.~Kurths, and N.~Wessel.
\newblock Classifying cardiac biosignals using ordinal pattern statistics and
  symbolic dynamics.
\newblock {\em Computers in Biology and Medicine}, 42(3):319--327, 2012.

\bibitem{11Nicolaou}
Nicoletta Nicolaou and Julius Georgiou.
\newblock Detection of epileptic electroencephalogram based on permutation
  entropy and support vector machines.
\newblock {\em Expert Systems with Applications}, 39(1):202--209, 2012.

\bibitem{09Ouyang}
Gaoxiang Ouyang, Xiaoli Li, Chuangyin Dang, and Douglas~A. Richards.
\newblock Deterministic dynamics of neural activity during absence seizures in
  rats.
\newblock {\em Phys. Rev. E}, 79:041146, Apr 2009.

\bibitem{12Faes}
Luca Faes, Giandomenico Nollo, and Alberto Porta.
\newblock Non-uniform multivariate embedding to assess the information transfer
  in cardiovascular and cardiorespiratory variability series.
\newblock {\em Computers in Biology and Medicine}, 42(3):290--297, 2012.

\bibitem{12Bian}
Chunhua Bian, Chang Qin, Qianli D.~Y. Ma, and Qinghong Shen.
\newblock Modified permutation-entropy analysis of heartbeat dynamics.
\newblock {\em Phys. Rev. E}, 85:021906, Feb 2012.

\bibitem{10Li}
Xiaoli Li and Gaoxiang Ouyang.
\newblock Estimating coupling direction between neuronal populations with
  permutation conditional mutual information.
\newblock {\em NeuroImage}, 52(2):497--507, 2010.

\bibitem{08Olofsen}
E.~Olofsen, J.~W. Sleigh, and A.~Dahan.
\newblock Permutation entropy of the electroencephalogram: a measure of
  anaesthetic drug effect.
\newblock {\em British Journal of Anaesthesia}, 101:810--821, 2008.

\bibitem{18Garland}
Joshua Garland, Tyler~R. Jones, Michael Neuder, Valerie Morris, James W.~C.
  White, and Elizabeth Bradley.
\newblock Anomaly detection in paleoclimate records using permutation entropy.
\newblock {\em Entropy}, 20(12), 2018.

\bibitem{12Ruiz}
Mar\'{\i}a del~Carmen Ruiz, Antonio Guillam\'on, and Antonio Gabald\'on.
\newblock A new approach to measure volatility in energy markets.
\newblock {\em Entropy}, 14(1):74--91, 2012.

\bibitem{08Zanin}
Massimiliano Zanin.
\newblock Forbidden patterns in financial time series.
\newblock {\em Chaos: An Interdisciplinary Journal of Nonlinear Science},
  18(1):013119, 2008.

\bibitem{09Zunino}
Luciano Zunino, Massimiliano Zanin, Benjamin~M. Tabak, Dar\'{\i}o~G. P\'erez,
  and Osvaldo~A. Rosso.
\newblock Forbidden patterns, permutation entropy and stock market
  inefficiency.
\newblock {\em Physica A: Statistical Mechanics and its Applications},
  388(14):2854--2864, 2009.

\bibitem{10Zunino}
Luciano Zunino, Massimiliano Zanin, Benjamin~M. Tabak, Dar\'{\i}o~G. P\'erez,
  and Osvaldo~A. Rosso.
\newblock Complexity-entropy causality plane: A useful approach to quantify the
  stock market inefficiency.
\newblock {\em Physica A: Statistical Mechanics and its Applications},
  389(9):1891--1901, 2010.

\bibitem{12Zunino}
Luciano Zunino, Aurelio {Fern\'andez Bariviera}, M.~Bel\'en Guercio, Lisana~B.
  Martinez, and Osvaldo~A. Rosso.
\newblock On the efficiency of sovereign bond markets.
\newblock {\em Physica A: Statistical Mechanics and its Applications},
  391(18):4342--4349, 2012.

\bibitem{11Zunino}
Luciano Zunino, Benjamin~M. Tabak, Francesco Serinaldi, Massimiliano Zanin,
  Dar\'{\i}o~G. P\'erez, and Osvaldo~A. Rosso.
\newblock Commodity predictability analysis with a permutation information
  theory approach.
\newblock {\em Physica A: Statistical Mechanics and its Applications},
  390(5):876--890, 2011.

\bibitem{19Pessa}
Arthur A.~B. Pessa and Haroldo~V. Ribeiro.
\newblock Characterizing stochastic time series with ordinal networks.
\newblock {\em Phys. Rev. E}, 100:042304, Oct 2019.

\bibitem{12Yan}
Ruqiang Yan, Yongbin Liu, and Robert~X. Gao.
\newblock Permutation entropy: A nonlinear statistical measure for status
  characterization of rotary machines.
\newblock {\em Mechanical Systems and Signal Processing}, 29:474--484, 2012.

\bibitem{08Matilla}
Mariano Matilla-Garc\'{\i}a and Manuel {Ruiz Mar\'{\i}n}.
\newblock A non-parametric independence test using permutation entropy.
\newblock {\em Journal of Econometrics}, 144(1):139--155, 2008.

\bibitem{20Olivares_1}
Felipe Olivares and Luciano Zunino.
\newblock Multiscale dynamics under the lens of permutation entropy.
\newblock {\em Physica A: Statistical Mechanics and its Applications},
  559:125081, 2020.

\bibitem{13Fadlallah}
Bilal Fadlallah, Badong Chen, Andreas Keil, and Jos\'e Pr\'{\i}ncipe.
\newblock Weighted-permutation entropy: A complexity measure for time series
  incorporating amplitude information.
\newblock {\em Phys. Rev. E}, 87:022911, Feb 2013.

\bibitem{09Liu}
Liu Xiao-Feng and Wang Yue.
\newblock Fine-grained permutation entropy as a measure of natural complexity
  for time series.
\newblock {\em Chinese Physics B}, 18(7):2690--2695, jul.

\bibitem{08Staniek}
Matth\"aus Staniek and Klaus Lehnertz.
\newblock Symbolic transfer entropy.
\newblock {\em Phys. Rev. Lett.}, 100:158101, Apr 2008.

\bibitem{16Amigo}
Jos\'e~M. Amig\'o, Roberto Monetti, Beata Graff, and Grzegorz Graff.
\newblock Computing algebraic transfer entropy and coupling directions via
  transcripts.
\newblock {\em Chaos: An Interdisciplinary Journal of Nonlinear Science},
  26(11):113115, 2016.

\bibitem{95Lopez}
R.~L\'pez-Ruiz, H.~L. Mancini, and X.~Calbet.
\newblock A statistical measure of complexity.
\newblock {\em Physics Letters A}, 209(5):321--326, 1995.

\bibitem{04Lamberti}
P.W Lamberti, M.T Martin, A~Plastino, and O.A Rosso.
\newblock Intensive entropic non-triviality measure.
\newblock {\em Physica A: Statistical Mechanics and its Applications},
  334(1):119--131, 2004.

\bibitem{08Bahraminasab}
A.~Bahraminasab, F.~Ghasemi, A.~Stefanovska, P.~V.~E. McClintock, and H.~Kantz.
\newblock Direction of coupling from phases of interacting oscillators: A
  permutation information approach.
\newblock {\em Phys. Rev. Lett.}, 100:084101, Feb 2008.

\bibitem{09Monetti}
Roberto Monetti, Wolfram Bunk, Thomas Aschenbrenner, and Ferdinand Jamitzky.
\newblock Characterizing synchronization in time series using information
  measures extracted from symbolic representations.
\newblock {\em Phys. Rev. E}, 79:046207, Apr 2009.

\bibitem{17Keller}
Karsten Keller, Teresa Mangold, Inga Stolz, and Jenna Werner.
\newblock Permutation entropy: New ideas and challenges.
\newblock {\em Entropy}, 19(3), 2017.

\bibitem{12Amigo}
Jos\'e~M. Amig\'o, Roberto Monetti, Thomas Aschenbrenner, and Wolfram Bunk.
\newblock Transcripts: An algebraic approach to coupled time series.
\newblock {\em Chaos: An Interdisciplinary Journal of Nonlinear Science},
  22(1):013105, 2012.

\bibitem{13Bunk}
W.~Bunk, J.~M. Amig\'o, T.~Aschenbrenner, and R.~Monetti.
\newblock A new perspective on transcripts by means of their matrix
  representation.
\newblock {\em The European Physical Journal Special Topics}, 222:363--381,
  2013.

\bibitem{17Zunino}
Luciano Zunino, Felipe Olivares, Felix Scholkmann, and Osvaldo~A. Rosso.
\newblock Permutation entropy based time series analysis: Equalities in the
  input signal can lead to false conclusions.
\newblock {\em Physics Letters A}, 381(22):1883--1892, 2017.

\bibitem{15Quintero}
C.~Quintero-Quiroz, S.~Pigolotti, M.~C. Torrent, and C.~Masoller.
\newblock Numerical and experimental study of the effects of noise on the
  permutation entropy.
\newblock {\em New Journal of Physics}, 17(9):093002, sep 2015.

\bibitem{07Amigo}
J.~M Amig{\'{o}}, S~Zambrano, and M.~A.~F Sanju{\'{a}}n.
\newblock True and false forbidden patterns in deterministic and random
  dynamics.
\newblock {\em Europhysics Letters}, 79(5):50001, jul 2007.

\bibitem{17McCullogh}
M.~McCullough, M.~Small, H.~H.~C. Iu, and T.~Stemler.
\newblock Multiscale ordinal network analysis of human cardiac dynamics.
\newblock {\em Philosophical Transactions of the Royal Society A: Mathematical,
  Physical and Engineering Sciences}, 375:20160292, 2017.

\bibitem{11Campanharo}
Andriana S. L.~O. Campanharo, M.~Irmak Sirer, R.~Dean Malmgren, Fernando~M.
  Ramos, and Lu\'{\i}s A.~Nunes. Amaral.
\newblock Duality between time series and networks.
\newblock {\em PLOS ONE}, 6(8):1--13, 08 2011.

\bibitem{17Zhang}
J.~Zhang, J.~Zhou, M.~Tang, H.~Guo, M.~Small, and Y.~Zou.
\newblock Constructing ordinal partition transition networks from multivariate
  time series.
\newblock {\em Scientific Reports}, 7:7795, 2017.

\bibitem{20Olivares}
F.~Olivares, M.~Zanin, L.~Zunino, and D.~G. P\'erez.
\newblock Contrasting chaotic with stochastic dynamics via ordinal transition
  networks.
\newblock {\em Chaos: An Interdisciplinary Journal of Nonlinear Science},
  30(6):063101, 2020.

\bibitem{13Monetti_1}
Roberto Monetti, Wolfram Bunk, Thomas Aschenbrenner, Stephan Springer, and
  Jos\'e~M. Amig\'o.
\newblock Information directionality in coupled time series using transcripts.
\newblock {\em Phys. Rev. E}, 88:022911, Aug 2013.

\bibitem{03Endres}
D.M. Endres and J.E. Schindelin.
\newblock A new metric for probability distributions.
\newblock {\em IEEE Transactions on Information Theory}, 49(7):1858--1860,
  2003.

\bibitem{21Pessa}
Arthur A.~B. Pessa and Haroldo~V. Ribeiro.
\newblock ordpy: A python package for data analysis with permutation entropy
  and ordinal network methods.
\newblock {\em Chaos: An Interdisciplinary Journal of Nonlinear Science},
  31(6):063110, 2021.

\bibitem{11Pedregosa}
Fabian Pedregosa, Ga{{\"e}}l Varoquaux, Alexandre Gramfort, Vincent Michel,
  Bertrand Thirion, Olivier Grisel, Mathieu Blondel, Peter Prettenhofer, Ron
  Weiss, Vincent Dubourg, Jake Vanderplas, Alexandre Passos, David Cournapeau,
  Matthieu Brucher, Matthieu Perrot, and {{\'E}}douard Duchesnay.
\newblock Scikit-learn: Machine learning in python.
\newblock {\em Journal of Machine Learning Research}, 12(85):2825--2830, 2011.

\bibitem{20Harris}
Charles~R. Harris, K.~Jarrod Millman, St\'efan~J. van~der Walt, Ralf Gommers,
  Pauli Virtanen, David Cournapeau, Eric Wieser, Julian Taylor, Sebastian Berg,
  Nathaniel~J. Smith, Robert Kern, Matti Picus, Stephan Hoyer, Marten~H. van
  Kerkwijk, Matthew Brett, Allan Haldane, Jaime~Fern\'andez del R\'{\i}o, Mark
  Wiebe, Pearu Peterson, Pierre G\'erard-Marchant, Kevin Sheppard, Tyler Reddy,
  Warren Weckesser, Hameer Abbasi, Christoph Gohlke, and Travis~E. Oliphant.
\newblock Array programming with numpy.
\newblock {\em Nature}, 585:357--362, 2020.

\bibitem{13Arbelaitz}
Olatz Arbelaitz, Ibai Gurrutxaga, Javier Muguerza, Jes\'us~M. P\'erez, and
  I\~{n}igo Perona.
\newblock An extensive comparative study of cluster validity indices.
\newblock {\em Pattern Recognition}, 46(1):243--256, 2013.

\bibitem{11Luxburg}
Ulrike von Luxburg, Robert~C. Williamson, and Isabelle Guyon.
\newblock Clustering: Science or art?
\newblock In Isabelle Guyon, Gideon Dror, Vincent Lemaire, Graham Taylor, and
  Daniel Silver, editors, {\em Proceedings of ICML Workshop on Unsupervised and
  Transfer Learning}, volume~27 of {\em Proceedings of Machine Learning
  Research}, pages 65--79, Bellevue, Washington, USA, 02 Jul 2012. PMLR.

\bibitem{14Lewis}
Michael Lewis.
\newblock {\em Flash Boys: A Wall Street Revolt}.
\newblock W. W. Norton \& Company, 2014.

\bibitem{15Pasquale}
Frank Pasquale.
\newblock {\em The Black Box Society}.
\newblock Harvard University Press, 2015.

\bibitem{21data}
Mario L\'opez.
\newblock Ordinal synchronization.
\newblock \url{https://gitlab.com/mario-lopez-perez/ordinal-synchronization},
  2021.
\newblock Accessed on October 15, 2021.

\end{thebibliography}

\end{document}